# Probing Short-Range Correlations in the van der Waals Magnet CrSBr by Small-Angle Neutron Scattering.

*Andrey Rybakov, Carla Boix-Constant, Diego Alba Venero, Herre S. J. van der Zant, Samuel Mañas-Valero\* and Eugenio Coronado*

A. Rybakov, C. Boix-Constant, S. Mañas-Valero, E. Coronado

Instituto de Ciencia Molecular (ICMol), Universitat de València, Catedrático José Beltrán 2, Paterna, 46980 Spain.
E-mail: samuel.manas@uv.es

D. Alba Venero

ISIS Neutron and Muon Facility, Science and Technology Facilities Council, Rutherford Appleton Laboratory, Chilton OX11 0QX, United Kingdom.

H. S. J. van der Zant, S. Mañas-Valero

Kavli Institute of Nanoscience, Delft University of Technology, Lorentzweg 1, 2628 CJ Delft, The Netherlands.
E-mail: S.ManasValero@tudelft.nl

The layered metamagnet CrSBr offers a rich interplay between magnetic, optical and electrical properties that can be extended down to the two-dimensional (2D) limit. Despite the extensive research regarding the long-range magnetic order in magnetic van der Waals materials, short-range correlations have been loosely investigated. By using Small-Angle Neutron Scattering (SANS) we show the formation of short-range magnetic regions in CrSBr with correlation lengths that increase upon cooling up to *ca.* 3 nm at the antiferromagnetic ordering temperature ($T_N \sim 140$ K). Interestingly, these ferromagnetic correlations start developing below 200 K, *i.e.*, well above $T_N$. Below $T_N$, these correlations rapidly decrease and are negligible at low-temperatures. The experimental results are well-reproduced by an effective spin Hamiltonian, which pinpoints that the short-range correlations in CrSBr are intrinsic to the monolayer limit, and discard the appearance of any frustrated phase in CrSBr at low-temperatures within our experimental window between 2 and 200 nm. Overall, our results are compatible with a spin freezing scenario of the magnetic fluctuations in CrSBr and highlight SANS as a powerful technique for characterizing the rich physical phenomenology beyond the long-range order paradigm offered by van der Waals magnets.

# 1. Introduction

Van der Waals (vdW) magnets are a broad family of layered materials that offer a versatile platform for addressing both fundamental questions in low-dimensional magnetism as well as applied developments in areas such spintronics, magnonics or data storage since the magnetic layers can be used as building blocks for the fabrication of magnetic vdW heterostructures.[1–3] Thanks to their rich chemical composition, different magnetic ground states can be found including conventional ferromagnetism (e.g., $Fe_3GeTe_2$ or $Cr_2Ge_2Te_6$)[4,5] or antiferromagnetism (e.g., $FePS_3$ or $NiPS_3$),[6] as well as more exotic behaviors such as frustrated magnetism (e.g., CeSiI),[7] quantum spin liquids (e.g., $RuCl_3$ or $1T-TaS_2$)[8–10] or different classes of metamagnetism (e.g., $CrI_3$, $CrPS_4$ or CrSBr).[11–13]. The structural features of these layered vdW materials —which typically shows strong exchange interactions within the layers but very weak interlayer interactions— provide a unique situation in which the emergence of long-range magnetic order is strongly dependent on the spin dimensionality.[14] In this regard, the role of short-range correlations coupling electronic and magnetic degrees of freedom is fundamental for understanding the properties of these materials. For instance, short-range correlations have been related to an enhancement of the thermoelectric properties,[15] the appearance of quantum phase transitions[16] and even to the origin of high-temperature superconductivity.[17] However, most of the experimental efforts regarding van der Waals magnets have focused on the long-range magnetic ordered phase, being the quantification of the short-range correlations relegated to a secondary place, likely due to the lack of proper experimental techniques able to its quantification.[14] Interestingly, thanks to the weak inter-layer interactions in vdW materials, all these properties can be studied down to the 2D limit and even controlled by proximity effects or twisting of the layers.[18–20]

In this work, we consider the layered magnetic semiconductor CrSBr, which has gained recent interest due to the interplay between its magnetic, optical, and electrical properties down to the 2D limit.[21–26] This material is formed by ferromagnetic layers that couple antiferromagnetically (**Figure 1.a**) undergoing a long-range magnetic ordering at $T_N \sim 140$ K.[21] In addition, upon the application of moderate magnetic fields, it is possible to reorient the spin of the layers (at 10 K: 0.6 T, 1 T and 2 T for fields applied along the *b*, *a* and *c* axes, respectively).[12] Previous muon and magneto-transport experiments have shown the existence of short-range correlations in CrSBr already below 200 K,[12,23,27] which are highly relevant regarding the light-matter interaction.[28–31] However, due to the emergence of long-range order at lower temperatures, it was not possible to quantify these short-range correlations. Here, we employ Small-Angle Neutron Scattering (SANS) in order to determine the correlation length of the short-range fluctuations, as well as its temperature dependence. In the context of magnetism, SANS is a technique able to resolve structures on a length scale between a few and a few hundred nanometers, thus being able to resolve magnetic short-range correlations, as already demonstrated,[32–35] while being insensitive to the long-range ones.[36–39] Our results show the appearance of short-range correlations below *ca.* 200 K, characterized by correlations lengths up to *ca.* 3 nm at $T_N$, and highlight SANS as a powerful technique for characterizing vdW magnets.

## 2. Results and discussion

Crystals of CrSBr are grown by a solid-state reaction (see **Methods**). CrSBr crystallizes in an orthorhombic space group, characterized by $\alpha = \beta = \gamma = 90°$, $a = 3.512$ Å, $b = 4.762$ Å and $c = 7.962$ Å,[12] being *c* related to the distance between the layers (**Figure 1.a**). SANS experiments are performed on a CrSBr powder sample in a transmission geometry (see **Methods**). As sketched in **Figure 1.b**, the incident neutron beam scatters in the sample and the diffuse magnetic scattering is recorded in a 2D detector. All 2D neutron data are shown in the **Supplementary Section 1**. Then, the SANS pattern is obtained by integrating over each azimuthal angle around the center of diffraction, with an experimental detection in the 0.004–0.7 Å$^{-1}$ Q range, being Q the wavevector. Magnetic fields are applied perpendicular to the incident neutron beam.

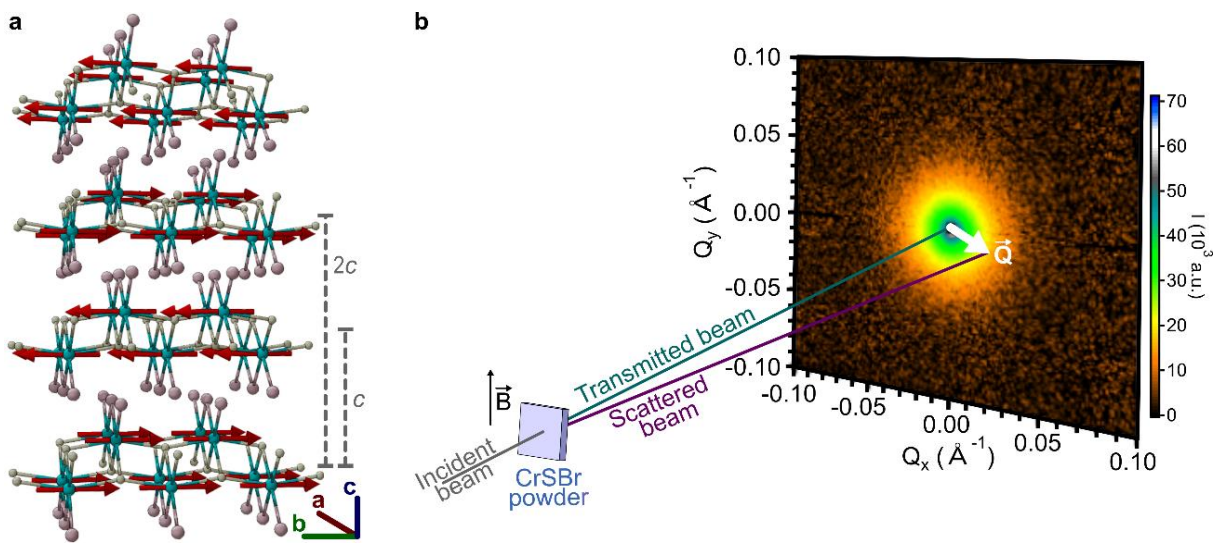

**Figure 1. CrSBr crystal structure and Small-Angle Neutron Scattering (SANS) experimental configuration.** a) Crystal structure of the layered vdW magnet CrSBr. The chromium, sulphur and bromine atoms are represented as cyan, yellow and pink balls, respectively. In the ordered state, the spins within each single layer couple ferromagnetically – spins represented as red arrows– pointing along the *b*-axis. The ferromagnetic layers couple antiferromagnetically among them along the *c*-axis. b) SANS experiment schematics in a transmission mode. The intensity of the scattered beam is recorded in a 2D detector (an example at 300 K is shown).

SANS patterns at different temperatures are shown in **Figure 2.a**. (the complete dataset is shown in **Supplementary Section 1**). At low momentum transfer, the pattern follows a $Q^{-4}$ dependence for the whole temperature range steaming from the powder nature of the sample, deviating at $Q \sim 0.025$ Å$^{-1}$ and flattening above $Q \sim 0.3$ Å$^{-1}$. Comparing the spectrum at high temperature (T = 300 K; orange color in **Figure 2.a**) and low temperature (T = 10 K; dark blue color in **Figure 2.a**), the only remarkable difference is the appearance of a Bragg peak at $Q \sim 0.407$ Å$^{-1}$ at low temperatures. This peak arises due to the antiferromagnetic interlayer coupling (**Figure 1.a**). In fact, taking into account the Bragg relation $Q = 2\pi/d$, d can be estimated to be 15.4 ± 0.5 Å, which matches well with a doubling of the cell along the *c*-axis. The thermal dependence of the intensity of this peak (**Figure 2.b**) corroborates its magnetic nature as it increases below $T_N$. In addition, application of an external field suppresses this peak since the spins are reoriented along the applied field and, therefore, the antiferromagnetic structure vanishes (**Figure 2.c**.).

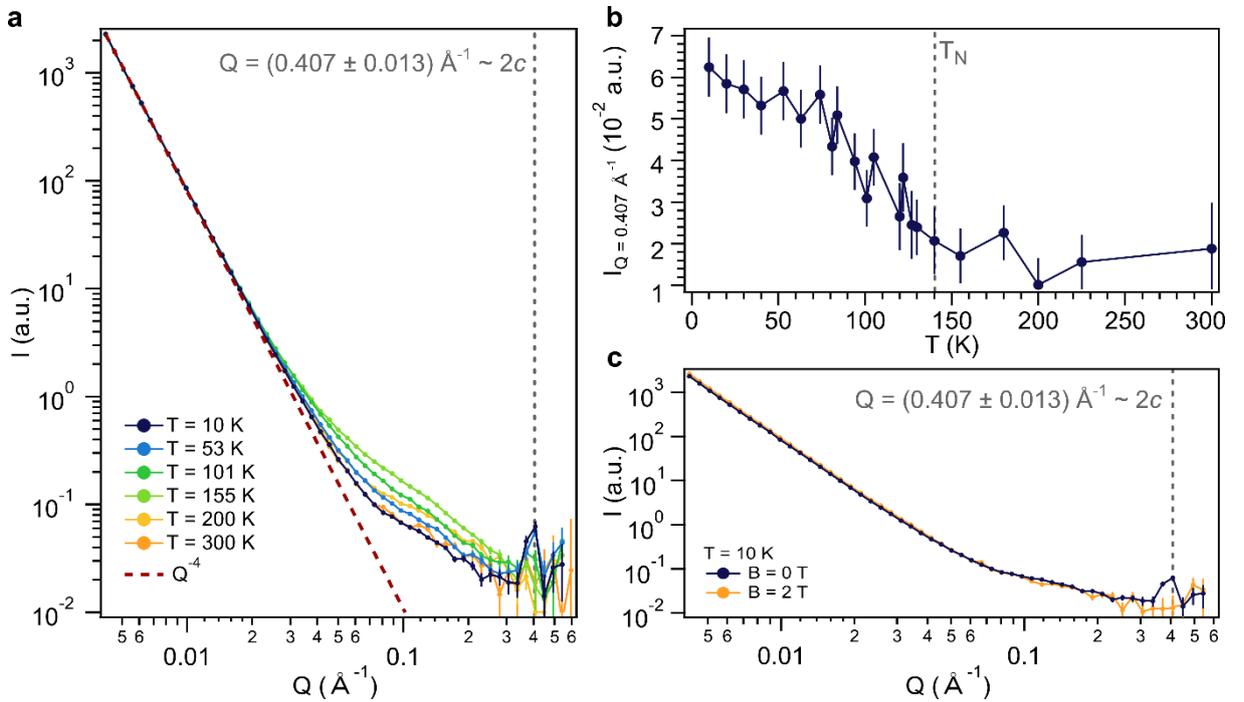

**Figure 2. Antiferromagnetic order in CrSBr probed by Small-Angle Neutron Scattering (SANS). a)** SANS pattern at selected temperatures (the whole dataset is presented in the **Supplementary Figure 1.a**). At low temperatures, a Bragg peak is observed at Q = 0.407 Å$^{-1}$ ~ 2*c* as a consequence of the antiferromagnetic order. **b)** Thermal dependence of the SANS intensity at Q = 0.407 Å$^{-1}$, where an enhancement of the signal is observed below the Néel temperature ($T_N \sim 140$ K). **c)** SANS spectra at T = 10 K. The peak at Q = 0.407 Å$^{-1}$ vanishes at high applied magnetics fields due to the spin-reorientation of the layers.

Both spectra at high and low temperatures (**Figure 2.a**) exhibit almost an identical Q dependence, except for the appearance of the Bragg peak, as expected for a long-range ordered antiferromagnet. Interestingly, the major variations in the spectra are observed in the range 0.025 – 0.3 Å$^{-1}$ at intermediate temperatures (**Figure 2.a** and **Supplementary Figure 2.a**). For a better visualization of these differences, we show in **Figure 3.a** the magnetic contribution of the SANS signal by subtracting the structural component (in our case, the spectra at 300 K in the high-temperature paramagnetic phase).[40] The complete Q-range is presented in the **Supplementary Figure 2.b**. While lowering the temperature, the SANS magnetic contribution is already detectable at 200 K, *i.e.,* 60 K above $T_N$, exhibiting its maximum at $T_N$ and starting to decrease upon further cooling down (**Figure 3.a**). An example of the SANS spectrum is shown in **Figure 3.b**. We determine the correlation length, ξ, based on an Ornstein-Zernike analysis, following a well stablished phenomenological method in SANS as previously reported[41–43] (see **Methods**). From the fit (**Figure 3.b**), ξ is estimated to be in the order of 3 nm at 140 K.

By performing similar fittings at different temperatures (**Supplementary Section 2**), the thermal dependence of ξ and the intensity scaling (related to the volume fraction of the correlated regions, assuming that the net magnetic moment is constant) is determined (**Figure 3.c**). ξ (blue dots in **Figure 3.c**) increases below 225 K up to a maximum around $T_N$. Below $T_N$, ξ diminishes and, at temperatures lower than 40 K, the magnetic signal in the studied region decreases very quickly due to the reduction in the volume fraction of the correlated region. This fact is better observed in the intensity scale (red squares in **Figure 3.c**). At high temperatures, $I_{OZ}(0)$ is almost zero, indicating that the magnetic contribution to the SANS signal is negligible. Below 200 K, $I_{OZ}(0)$ increases and reaches its maximum at $T_N$. Below $T_N$, $I_{OZ}(0)$ decreases rapidly (by a factor of 3) in the 80 K - 140 K range and, finally, is suppressed below 40 K (thus, a short-range frustrated magnetic state in CrSBr, as it could be speculated based on the magnetic exchange interactions,[23] is not likely to occur at very low-temperatures). Overall, the observed magnetic behavior is in agreement with a spin-freezing scenario occurring in CrSBr, as previously suggested by magnetization and muon experiments,[23] where only long-range interactions become relevant at low-temperatures and, therefore, are not detected in our SANS signal. In contrast with previous observations, where a slowing down of the magnetic fluctuations was observed below *ca.* 100 K, we observe a rapid suppression of these short-range correlations just below $T_N$, becoming negligible below *ca.* 40 K. Summarizing, we observe correlated regions with a net magnetic moment well above the ordering temperature, that increase while approaching the ordering temperature and are characterized by a correlation length in the order of 3 nm at $T_N$. By cooling below $T_N$, these magnetic fluctuations decrease, as observed by the diminishment of the intensity scaling, $I_{OZ}(0)$.

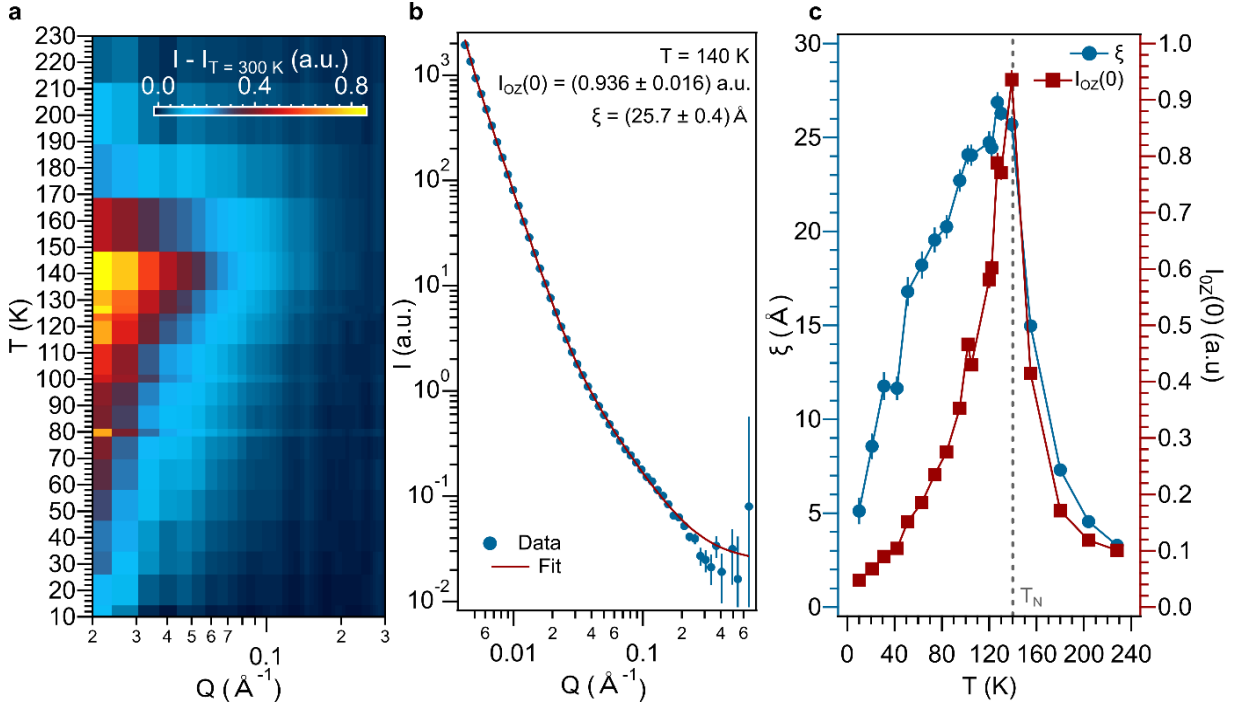

**Figure 3. Thermal dependence of the short-range correlations in CrSBr probed by SANS. a)** Thermal dependence of the magnetic contribution of the SANS signal after removing the structural component ($I_{T = 300 K}$). **b)** SANS signal at T = 140 K fitted following an Ornstein-Zernike law, yielding to an estimate of the volume fraction, $I_{OZ}$, and correlation length, $\xi$ (see text for details). **c)** Thermal dependence of the volume fraction and correlation length. The Néel temperature is marked as a grey dashed line. The complete set of fittings is presented in the **Supplementary Section 2**.

As a secondary fingerprint of the short-range correlations, we consider the role of applying an external magnetic field. In **Figure 4.a**, we show the magnetic contribution to the SANS signal at $T_N$ upon the application of different external magnetic fields, observing a suppression of the signal as the field is increased. By performing the same analysis as discussed above, we determine $\xi$ and intensity scaling (**Figure 4.b**), which diminish as the applied field is increased. A similar field dependence is observed at 150 K (**Supplementary Section 2**). On the contrary, no field dependence in the region between 0.02-0.3 Å$^{-1}$ is observed at 10 K (**Supplementary Section 2**), in agreement with a scenario with only long-range order at low temperatures and, therefore, not sensitive by SANS.

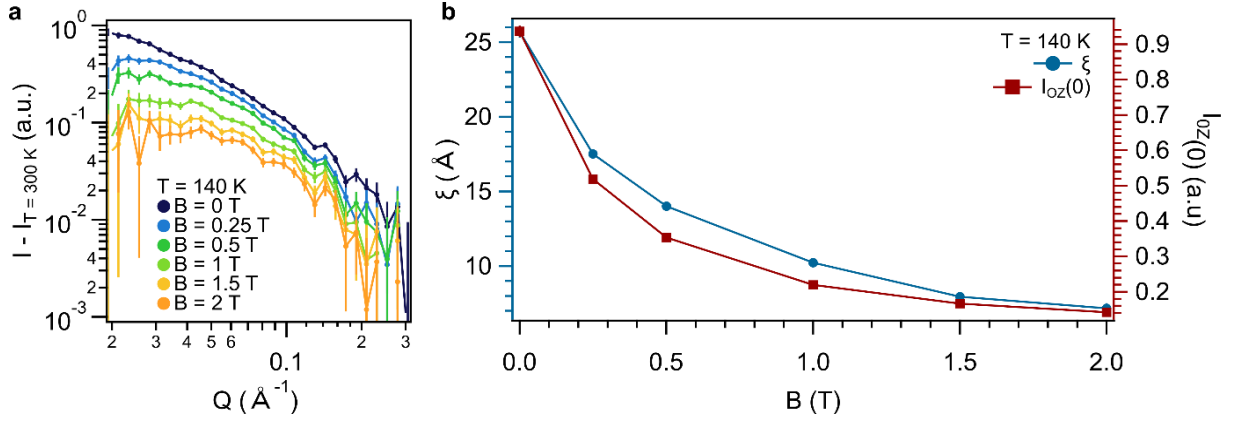

**Figure 4.- Field dependence of the short-range correlations in CrSBr at T = 140 K. a)** Magnetic contribution of the SANS signal at different applied magnetic fields after removing the structural component ($I_{T=300\,K}$). **b)** Field dependence of the volume fraction and correlation length. The complete set of fittings is presented in the **Supplementary Section 2**.

For modelling the short-range correlations in bulk CrSBr, we employ the following effective spin Hamiltonian:

$$H = J_{ij}\sum_{i\neq j}\vec{S}_i\cdot\vec{S}_j + \sum_{i\neq j}\vec{D}_{ij}\cdot\vec{S}_i\times\vec{S}_j + \sum_i\vec{S}_iA_i\vec{S}_i + H_{dd}.$$

The first term describes the bilinear exchange interactions, where $J_{ij}$ is an isotropic exchange parameter up to seven neighbor's order in the layers[44] (extracted from inelastic neutron scattering experiments) and up to second neighbor's order between the layers (taken from first-principles calculations).[45] The second term takes into account the antisymmetric anisotropic Dzyaloshinskii–Moriya interaction (DMI) that are allowed for some bonds of the CrsBr based on its symmetry (following the first-principles calculations).[45] The third term reflects the triaxial on-site anisotropy with the values adopted from microwave absorption spectroscopy measurements with $A_i$ being a diagonal matrix .[46] The last term accounts for magnetic dipole-dipole interactions. For every temperature and magnetic field, we calculate the spin-spin correlation function from the result of the Monte-Carlo sampling as implemented in the VAMPIRE computational package.[47] The calculated correlation function for bulk exhibits an exponential decay (**Supplementary Section 3**) and is separated into the interlayer and intralayer contribution due to the relatively small exchange between the layers of CrSBr (in the range of $10^{-3}$ meV) as compared to the intralayer ones (in the range of 1 meV), indicating that the 2D intralayer short-range correlations are the main ones in this material.[44] The intralayer term dominates the exponential decay at each temperature while the interlayer contribution remains close to zero for all temperatures and exhibits little or none temperature and field dependence, thus pointing towards the appearance of long-range interlayer antiferromagnetic interactions only below $T_N$. By fitting the correlation function (see **Methods**), we obtain the computed correlation length as a function of temperature (**Figure 5.a**) and magnetic field (**Figure 5.b**), reproducing well our experimental observations. To further illustrate the key role of the intralayer interactions, we compute as well the thermal dependency of the correlation length for a monolayer (blue squares in **Figure 5.a**). Compared to the bulk case (red dots in **Figure 5.a**), the bulk thermal dependency of the correlation length mimics the behavior of the monolayer, reflecting the major role of the intralayer interactions and highlighting that the short-range correlations in CrSBr arise from the monolayer.

Regarding the magnetic field dependence at T = 140 K (**Figure 5.b**), we apply the field along the three main crystallographic axes in three separate simulations. Fields along the medium (*a*) or hard (*c*) axis result in a delayed decay of $\xi$ if compared with the field aligned along the easy (*b*) axis (**Supplementary Section 3**). Since, experimentally, in a powder sample all directions are present, we consider as well the average trend. Overall, the dynamic correlations are suppressed as the field is increased due to the appearance of long-range order, in agreement with the experimental result (**Figure 4.b**). Finally, we consider the underlying spin textures appearing in CrSBr while cooling down (**Figure 5.c-e**). First, above $T_N$, the calculated $\xi$ increases as the sample is cooled down, which corresponds to the formation of correlated domains (**Figure 5.c**), while the long-range order is not present. It reaches the maximum around the phase transition temperature (**Figure 5.d**) and, then, it decreases and saturates at low temperatures, where the long-range order dominates (**Figure 5.e**). In the region below the phase transition, long range order is established and, therefore, the dynamical correlations are effectively suppressed, as evidenced from the results of the simulations.

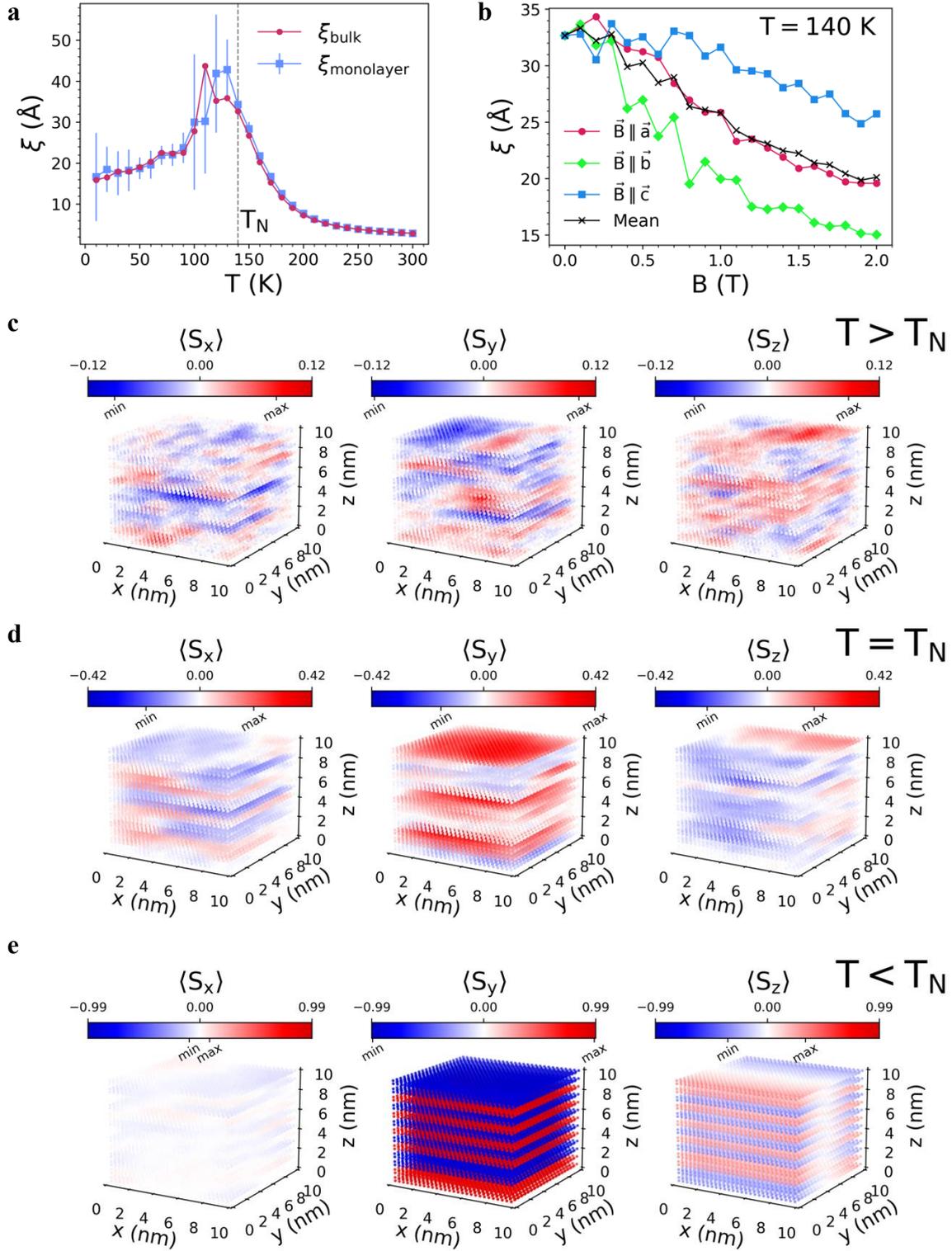

**Figure 5.- Simulated thermal dependence of the microscopic magnetization in CrSBr. a)** Temperature dependence of the correlation length for the intralayer pairs of bulk sample (red circles) and for monolayer sample (blue squares). The error bars for the monolayer indicate one standard deviation. **b)** Field dependence of the correlation length at 140 K for the intralayer pairs of bulk sample. **c-e)** Equilibrium spin distribution of the microscopic magnetic moments in the simulated bulk sample in the paramagnetic phase (T = 170 K, panel **c**), at the transition temperature (T = 140 K, panel **d**) and in the ordered phase (T = 20 K, panel **e**). Colors indicate

the value of the corresponding magnetization component from minimal (blue) to maximal (red) value, normalization is the same for all components. Minimum and maximum values of each individual component are noted below the color bars. The *a* crystallographic axis of the material is oriented along *x*; *b* - along *y*; *c* - along *z*.

## 3. Conclusion

In this work we have probed the role of short-range correlations in the van der Waals magnet CrSBr by Small-Angle Neutron Scattering experiments. Our experimental results quantify the short-range magnetic fluctuations in CrSBr, which are characterized by a correlation length in the order of 3 nm at the ordering temperature, $T_N$, and confirm the antiferromagnetic ordering below 140 K as well as the absence of frustrated magnetic states at low-temperatures. These correlations exhibit an interesting thermal dependence since they are already present below 200 K —that is, well above $T_N$ —, they exhibit a maximum at $T_N$ and, then, decrease rapidly while cooling down, being absent at low-temperatures. In accordance, the application of an external field suppresses these fluctuations. In addition, these experimental observations are well reproduced by a theoretical model based on an effective spin Hamiltonian, highlighting that the appearance of short-range correlations are intrinsic to the monolayer limit. Overall, our results are in accordance with a spin-freezing scenario in CrSBr, where the magnetic fluctuations cease while cooling down, and highlight SANS as an optimal technique for characterizing the rich physical phenomenology occurring in van der Waals magnets beyond the conventional long-range order picture.

## 4. Methods

*Crystal growth:* Crystals of CrSBr are grown by solid-state techniques, as previously reported by some of us.[12] The crystal structure is verified by powder and single-crystal X-ray diffraction together with the elemental composition by energy-dispersive X-ray spectroscopy (EDS).

*SANS measurements:* SANS experiments are performed at the Larmor instrument (ISIS neutron and muon source, United Kingdom) on 115 mg of CrSBr powder sample. The coherence length of the neutron beam is tens of micrometers, but at least higher than 57 nm, as experimentally probed.[42] The incident neutron beam scatters in the sample in a transmission geometry, being the diffuse magnetic scattering recorded in a 2D detector (sample-detector distance is 4 m, with wavelengths from 0.9 to 13 Å). The signal is integrated over each azimuthal angle around the center of diffraction, with an experimental detection in the 0.004–0.7 Å$^{-1}$ Q range, being Q the wavevector. SANS spectra are recorded as a function of temperature (10 K – 300 K) and magnetic field (0 T – 2 T), being the magnetic field applied perpendicular to the incident neutron beam. No significant thermal dependence of the SANS pattern is expected unless some structural or magnetic inhomogeneities in the range between 2-200 nm –that is, within our experimental sensitivity– start developing.[41] The neutron data reduction (including conversion of the Time of Flight and corrections for accounting the background scattering and transmission) is performed using the Mantid software.[48] Data fits are done with SasView application (http://www.sasview.org/, using dQ Data instrumental smearing). The data is fitted considering two different approaches. In the first approach (**Supplementary Section 2.a**), we fit the spectra at 300 K (high-temperature paramagnetic phase) to a power law, being $I_{T=300K}(Q) = \frac{I_{P,300K}}{Q^{4-n,300K}} + B_{300K}$, where $I_P$ is a Porod scale term (a particularly common dependence which main contributions are the divergence of the incident neutron beam as well as the Porod scattering from interfaces such as powder grain surfaces and grain boundaries, applicable to our powder measurement),[35] n is an exponent and B is a Q-independent background constant. Then, we employ the relationship $I_{T\neq 300K}(Q) = \frac{I_{OZ}(0)}{1+(\xi Q)^2} + \frac{I_{P,300K}}{Q^{4-n,300K}} + B$, where $I_{OZ}(0)$ is the Ornstein-Zernike intensity scaling and ξ is the correlation length. In the second approach (**Supplementary Section 2.b**), we consider $I(Q) - I_{300K}(Q) = \frac{I_{OZ}(0)}{1+(\xi Q)^2} + \frac{I_P}{Q^4} + B$. We note that both approaches are compatible between them, although the second one yields to a correlation length with larger error bars at low temperatures, that arise from an overparametrized fitting due to the absence of correlations within our experimental window range resolution. Despite being ξ constant below $T_N$, the absence of correlations is accounted by the suppression of $I_{OZ}(0)$. Fits shown in the main text are obtained with the first approach. All fitted data following both approaches are shown in the **Supplementary Section 2**.

*Computational details:* Spin-spin correlation function is calculated from the result of the Monte-Carlo sampling as implemented in the VAMPIRE computational package.[47] The computation is performed considering a 10x10x10 (nm) sample with open boundary conditions. The lateral size of the simulated sample is chosen in a balance of covering the measured range of the correlation length and the computational cost of the simulation. $10^5$ Monte-Carlo steps are computed for each temperature and field with first $3 \cdot 10^4$ reserved for the thermalization. From the microscopic spin distribution, the spin-spin correlation function $g(r_i, r_j) = \langle \vec{S}_i \vec{S}_j \rangle - \langle \vec{S}_i \rangle \langle \vec{S}_j \rangle$ is computed for each temperature, where $\langle ... \rangle$ denotes the canonical ensemble average. Next, the distance-dependent function $g(r_i - r_j)$ is computed by averaging over the pairs with the same distances. Finally, the correlation length is computed assuming the exponential decay of the spin-spin correlation function over distance. In particular, we consider the exponential fit $A \cdot e^{-r/\xi}$ for the intralayer part of the correlation function, where A is the amplitude, r the distance and ξ the correlation length. In order to account for the surface effects only the pairs with the distance smaller than 5nm are considered for the fit. For constructing the spin Hamiltonian, the intralayer isotropic exchange parameters were extracted from inelastic neutron scattering experiments.[44] The interlayer isotropic and intralayer DMI parameters are taken from first-principles calculations.[45] The on-site anisotropy values are adopted from the results of the microwave absorption spectroscopy measurements.[46] We remark that the notation of the spin Hamiltonians differs between sources; thus, the values are converted during the construction of the Hamiltonian (see **Supplementary Table 1** for details). After the construction of the Hamiltonian all parameters are scaled by a factor 1.5 in order to match the experimental value of $T_N$. In the calculations, the magnetic alignment of the layers does not follow the sequence expected for a pure A-type antiferromagnet. Still, the total balance of the spin-up and spin-down layers is preserved (thus out of 13 layers calculated in **Figure 5.c**, 6 are spin-up and 7 are spin-down). This result is not unexpected, as the interlayer exchange parameters are *ca.* 1000 times smaller than the intralayer ones. The intralayer correlation length saturates at the value ~2 nm below the phase transition, which is comparable with the result of the measurements. However, it is less stable in the vicinity of the phase transition. This suggests that the description of the equilibrium properties near the phase transition with the Monte-Carlo simulations may require considerably more iteration steps, which is not feasible from the computational point of view for the bulk.[49] For the monolayer case, we start the Monte-Carlo sampling from 20 different random initial spin configurations and estimate the mean value and standard deviation of the correlation length for each temperature.

**Supporting Information**

Supporting Information is available from the author.

**Acknowledgements**

The authors acknowledge the financial support from the European Union (ERC AdG Mol-2D 788222, FET OPEN SINFONIA 964396), the Spanish MCIN (2D-HETEROS PID2020-117152RB-100, co-financed by FEDER, and Excellence Unit "María de Maeztu" CEX2019-000919-M) and the Generalitat Valenciana (PROMETEO Program, PO FEDER Program IDIFEDER/2021/078, a Ph.D fellowship to C.B.-C and a Grisolia Ph.D fellowship to A. R. (GRISOLIAP/2021/038). This study forms part of the Advanced Materials program and was supported by MCIN with funding from European Union NextGenerationEU (PRTR-C17.I1) and by Generalitat Valenciana. S.M.-V. acknowledges the support from the European

Commission for a Marie Sklodowska–Curie individual fellowship No. 101103355 - SPIN-2D-LIGHT. We thank Á. López-Muñoz for his constant technical support and fundamental insights. We acknowledge the ISIS Neutron and Muon Facility for the provision of beamtime (RB2369003). This work benefited from the use of the SasView application, originally developed under NSF award DMR-0520547. SasView contains code developed with funding from the European Union's Horizon 2020 research and innovation programme under the SINE2020 project, grant agreement No 654000. Computations were performed on the HPC systems Cobra and Raven at the Max Planck Computing and Data Facility.

**ToC figure**

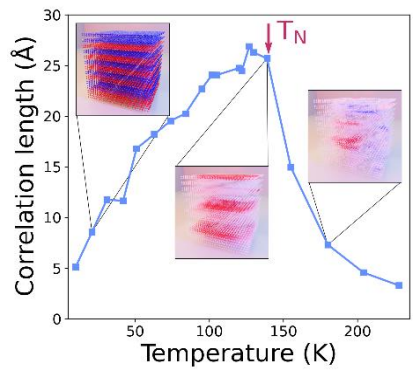

The correlation length of the layered magnetic semiconductor CrSBr is measured by Small-Angle Neutron Scattering. The ferromagnetic correlations start developing below 200 K, *i.e.*, well above $T_N$ (140 K), with a length of *ca.* 3 nm at $T_N$. The experimental results are well-reproduced by an effective spin Hamiltonian, which pinpoints that the short-range correlations are intrinsic to the monolayer limit.

# Supporting Information

**Probing Short-Range Correlations in the van der Waals Magnet CrSBr by Small-Angle Neutron Scattering.**

*Andrey Rybakov, Carla Boix-Constant, Diego Alba Venero, Herre S. J. van der Zant, Samuel Mañas-Valero* and Eugenio Coronado*

A. Rybakov, C. Boix-Constant, S. Mañas-Valero, E. Coronado
Instituto de Ciencia Molecular (ICMol), Universitat de València, Catedrático José Beltrán 2, Paterna, 46980 Spain.
E-mail: samuel.manas@uv.es

D. Alba Venero
ISIS Neutron and Muon Facility, Science and Technology Facilities Council, Rutherford Appleton Laboratory, Chilton OX11 0QX, United Kingdom.

H. S. J. van der Zant, S. Mañas-Valero
Kavli Institute of Nanoscience, Delft University of Technology, Lorentzweg 1, 2628 CJ Delft, The Netherlands.
E-mail: S.ManasValero@tudelft.nl

This file contains the **Supplementary Figures 1-16** and **Supplementary Table 1**.

**Contents:**

1. **Experimental data.**

    a. **SANS signal.**

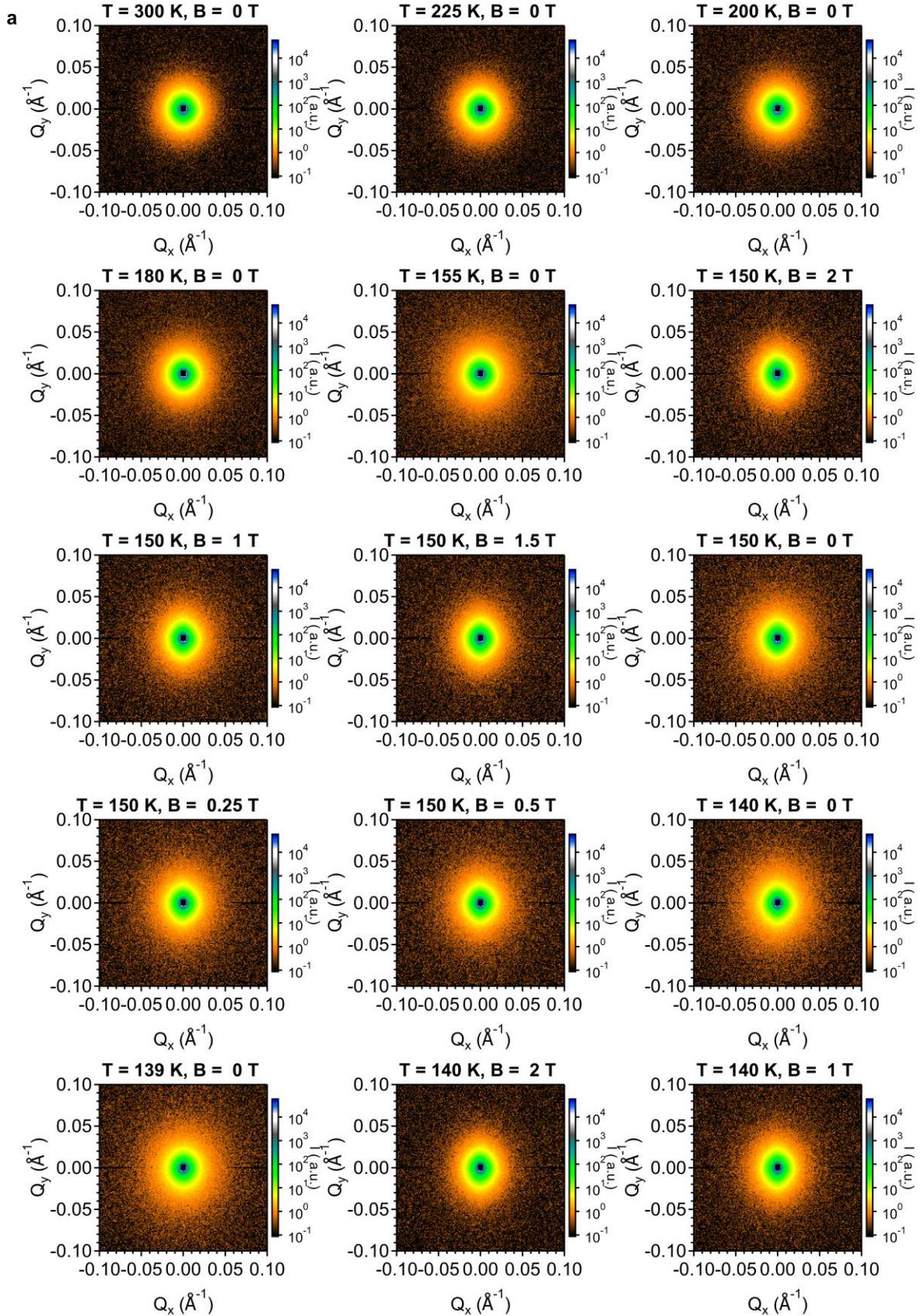

**Supplementary Figure 1.-** SANS signal at different temperatures and magnetic fields.

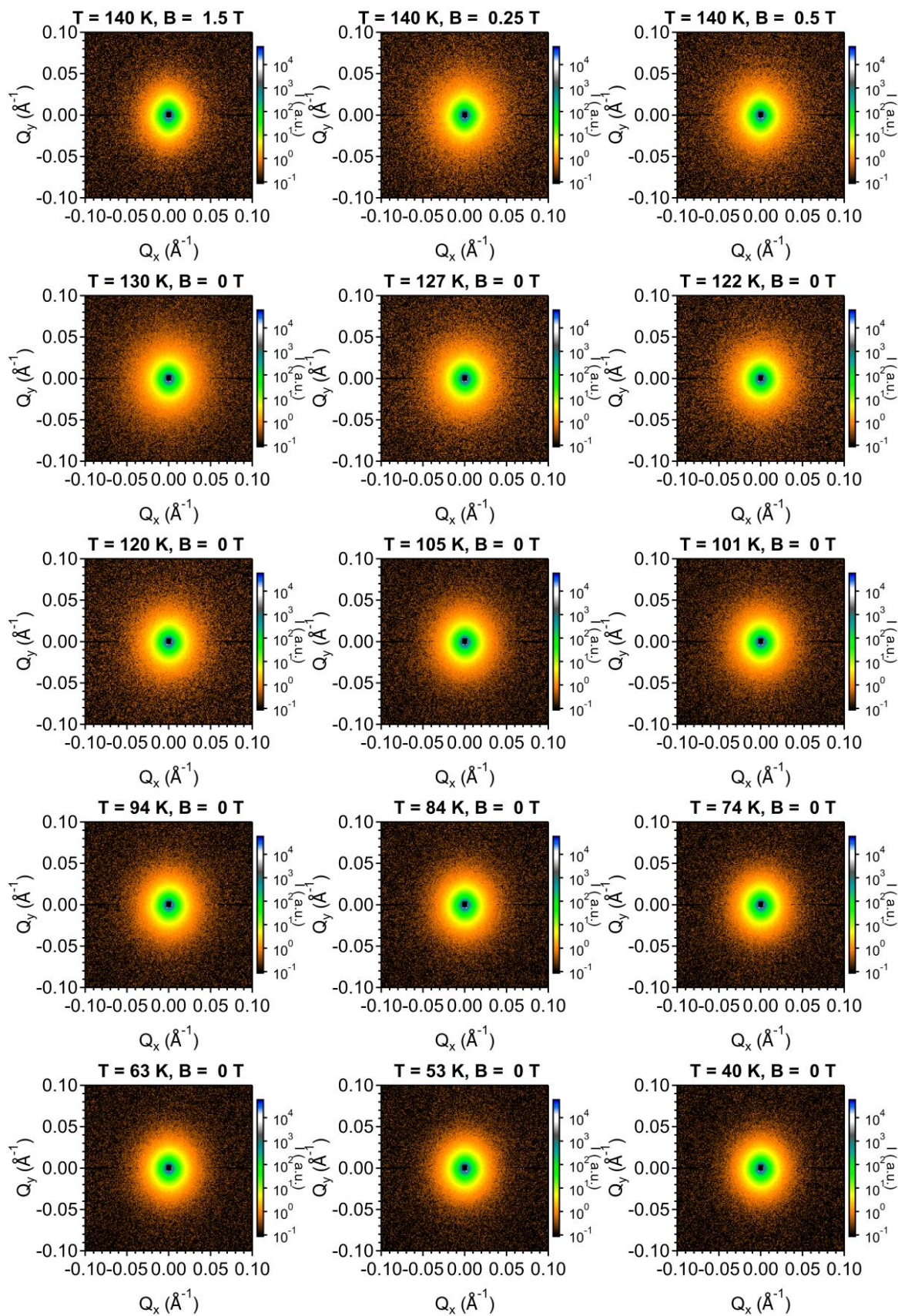

**Supplementary Figure 1.- (continues from previous page)** SANS signal at different temperatures and magnetic fields.

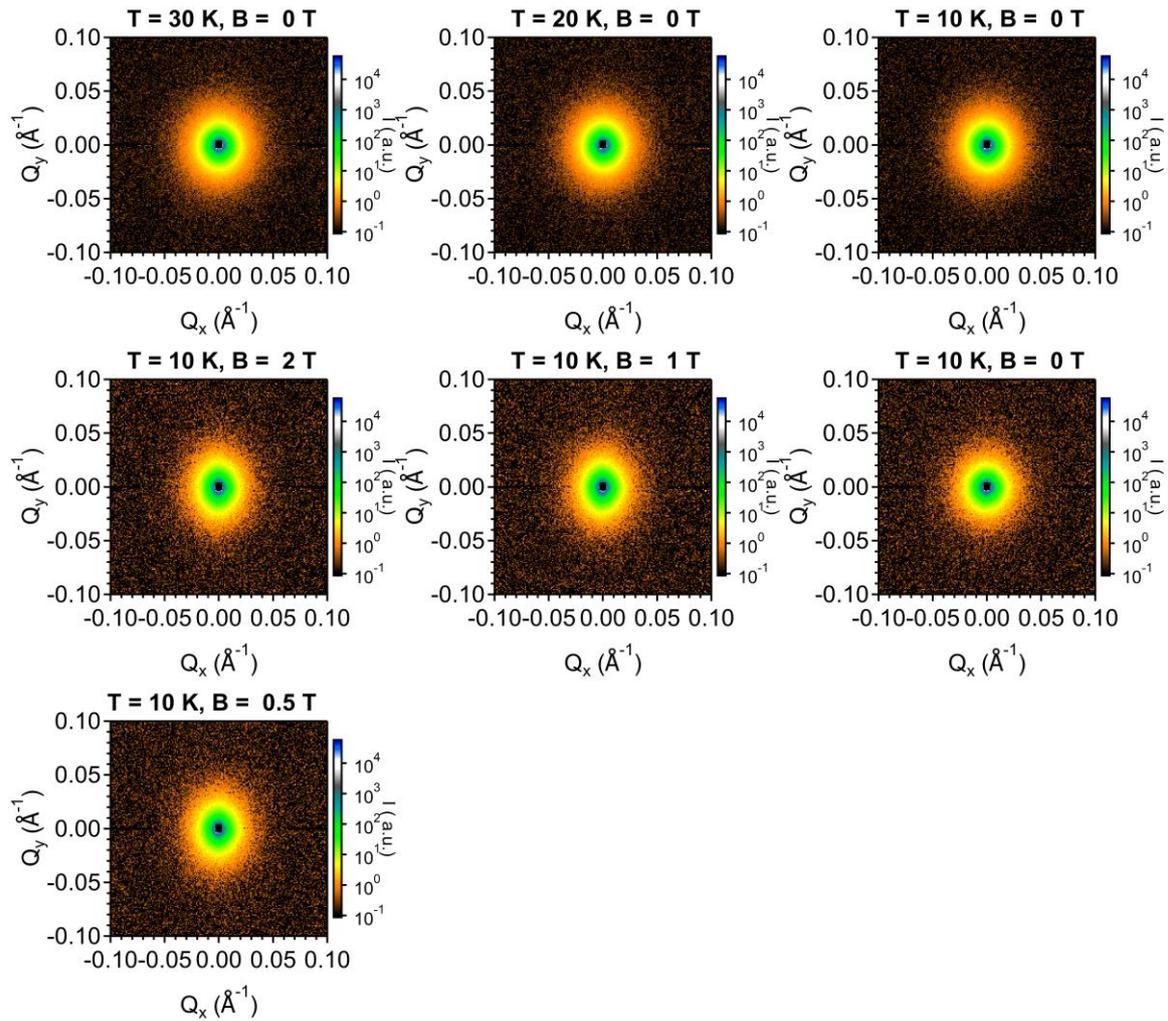

**Supplementary Figure 1.- (continues from previous page)** SANS signal at different temperatures and magnetic fields.

b. **Integrated SANS signal (temperature dependence).**

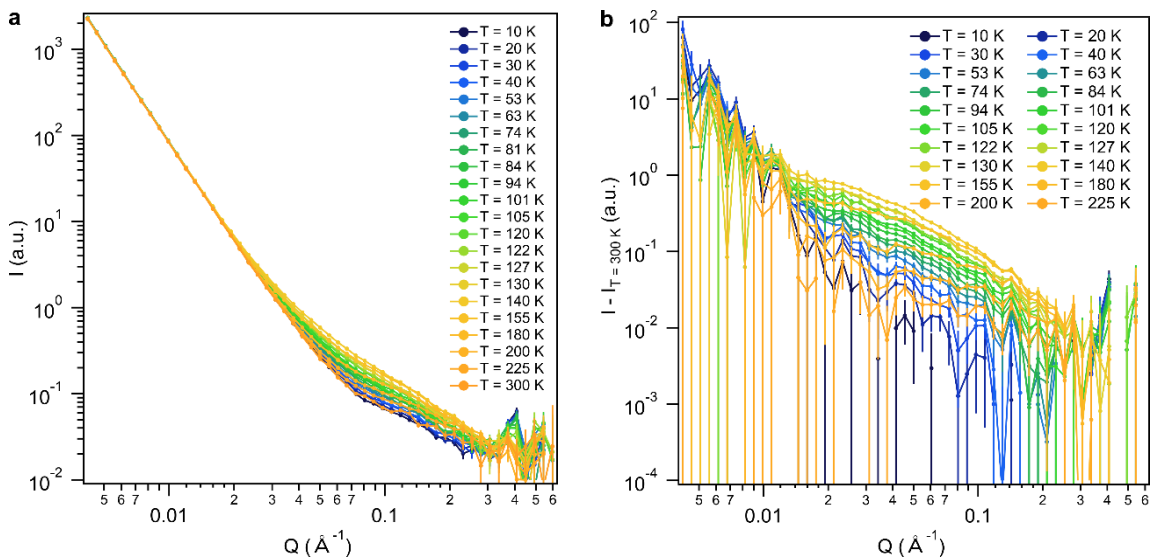

**Supplementary Figure 2.- a)** SANS signal at different temperatures. **b)** Magnetic contribution of the SANS signal after removing the structural component ($I_{T = 300 K}$) at different temperatures.

## c. Integrated SANS signal (field dependence).

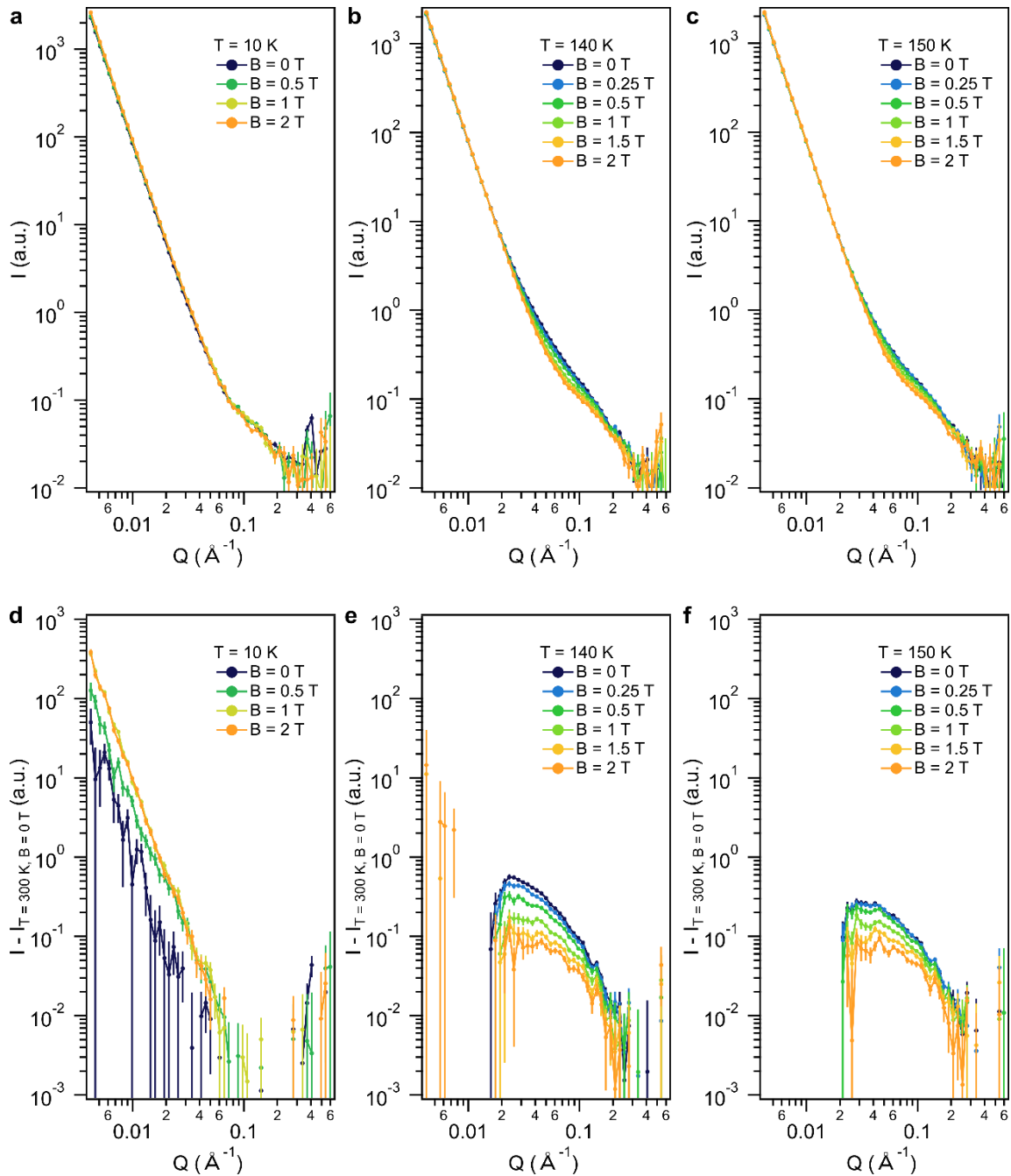

**Supplementary Figure 3.-** SANS signal (**a-c**) and magnetic contribution of the SANS signal (**d-f**) at different temperatures and magnetic fields.

**2. Data analysis.**

The data is fitted considering two different approaches. In the first approach (**Supplementary Section 2.a**), we fit the spectra at 300 K (high-temperature paramagnetic phase) to a power law, being $I_{T=300K}(Q) = \frac{I_{P,300K}}{Q^{4-n,300K}} + B_{300K}$, where $I_P$ is a Porod scale term, n is an exponent and B is a Q-independent background constant. Then, we employ the relationship $I_{T\neq 300K}(Q) = \frac{I_{OZ}(0)}{1+(\xi Q)^2} + \frac{I_{P,300K}}{Q^{4-n,300K}} + B$, where $I_{OZ}(0)$ is the Ornstein-Zernike intensity scaling and $\xi$ is the correlation length. In the second approach (**Supplementary Section 2.b**), we consider $I(Q) - I_{300K}(Q) = \frac{I_{OZ}(0)}{1+(\xi Q)^2} + \frac{I_P}{Q^4} + B$. We note that both approaches are compatible between them, although the second one yields to a correlation length with larger error bars at low temperatures, that arise from an overparametrized fitting due to the absence of correlations within our experimental window range resolution. Despite being $\xi$ constant below $T_N$, the absence of correlations is accounted by the suppression of $I_{OZ}(0)$. Fits shown in the main text are obtained with the first approach.

### a. I(T) analysis.

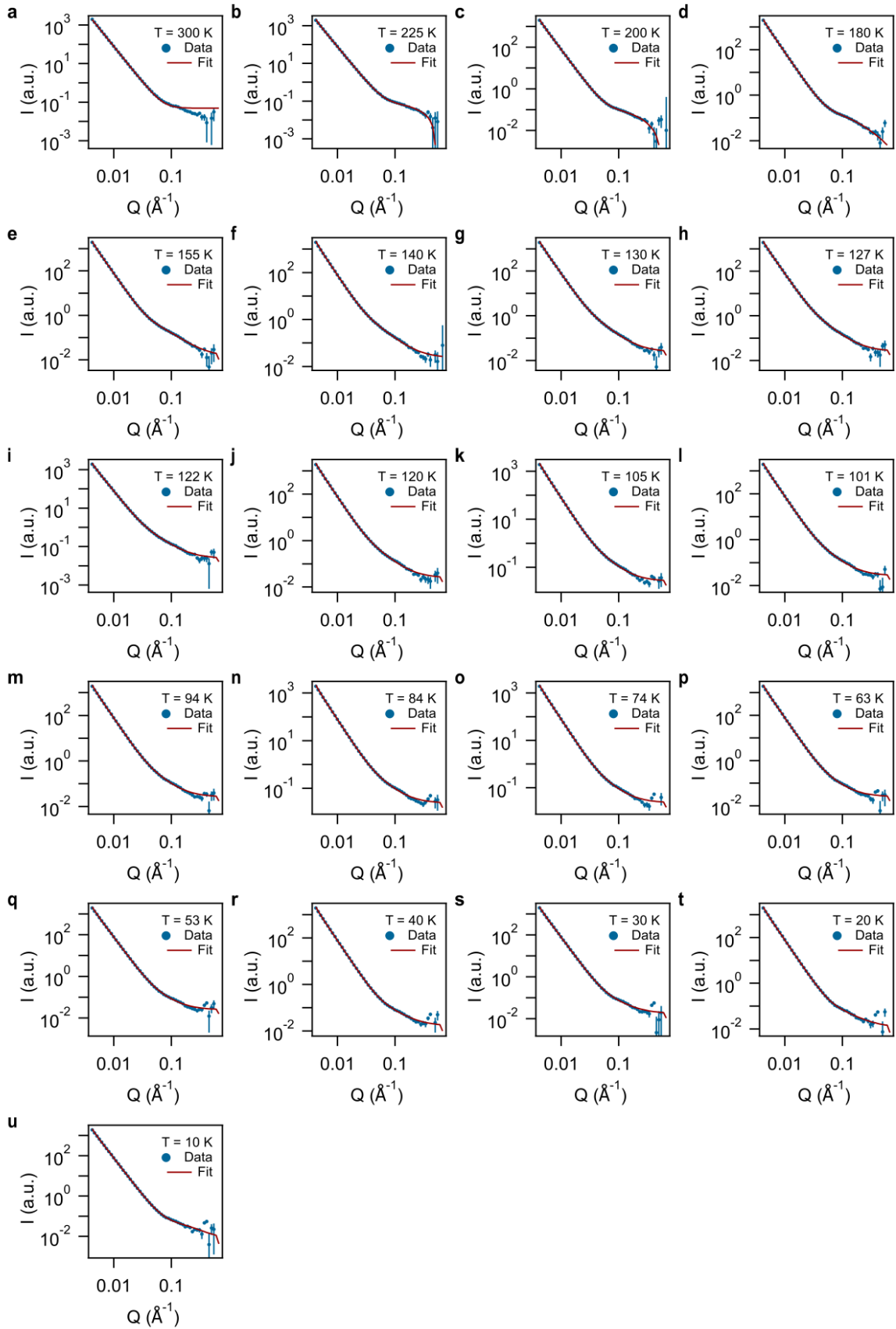

**Supplementary Figure 4.-** Fitted SANS signal at different temperatures (see Methods).

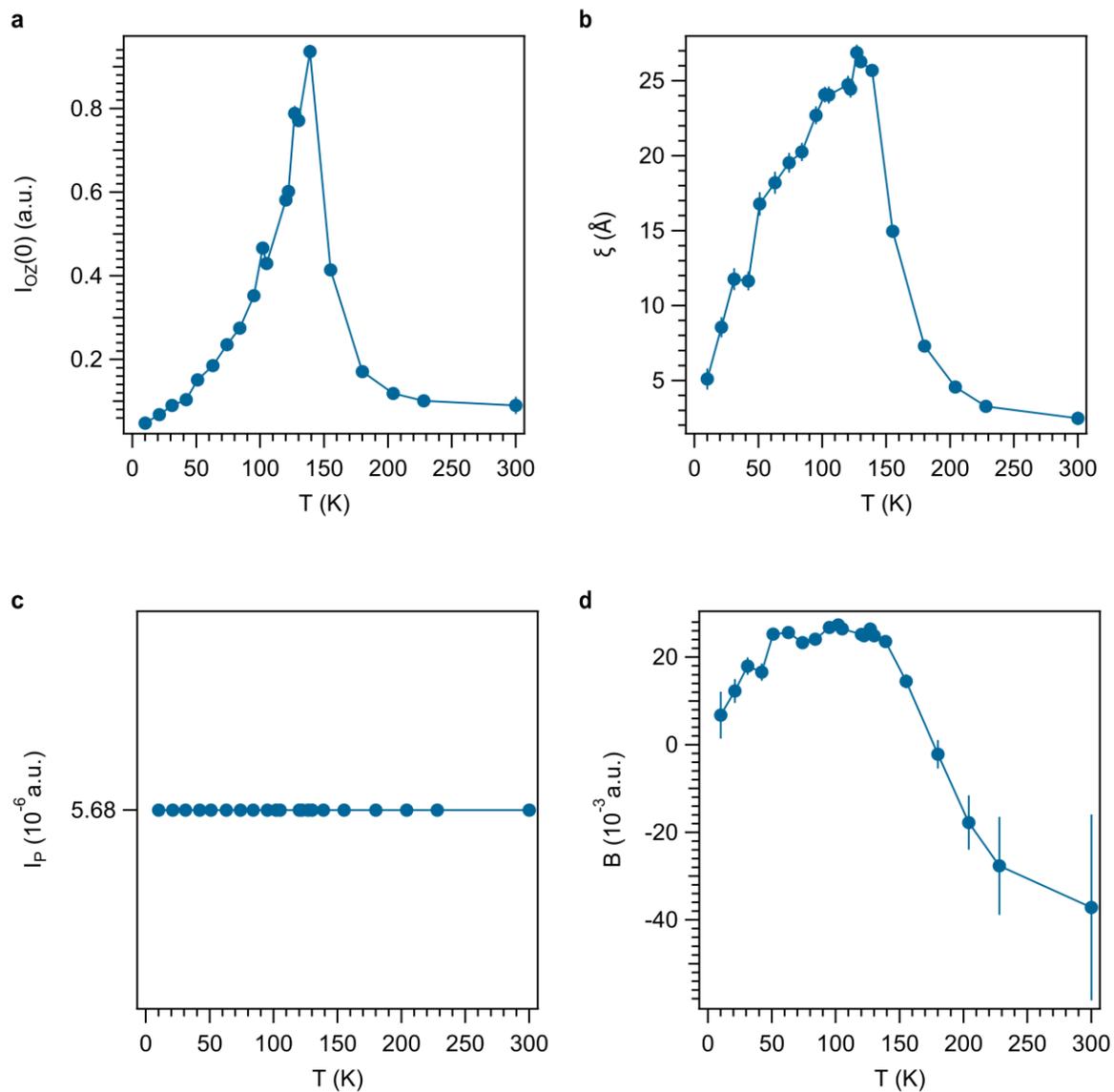

**Supplementary Figure 5.-** Thermal evolution of the intensity scale **(a)**, correlation length **(b)**, Porod scale factor **(c)** and background term **(d)** obtained from the fits shown in the **Supplementary Figure 4**.

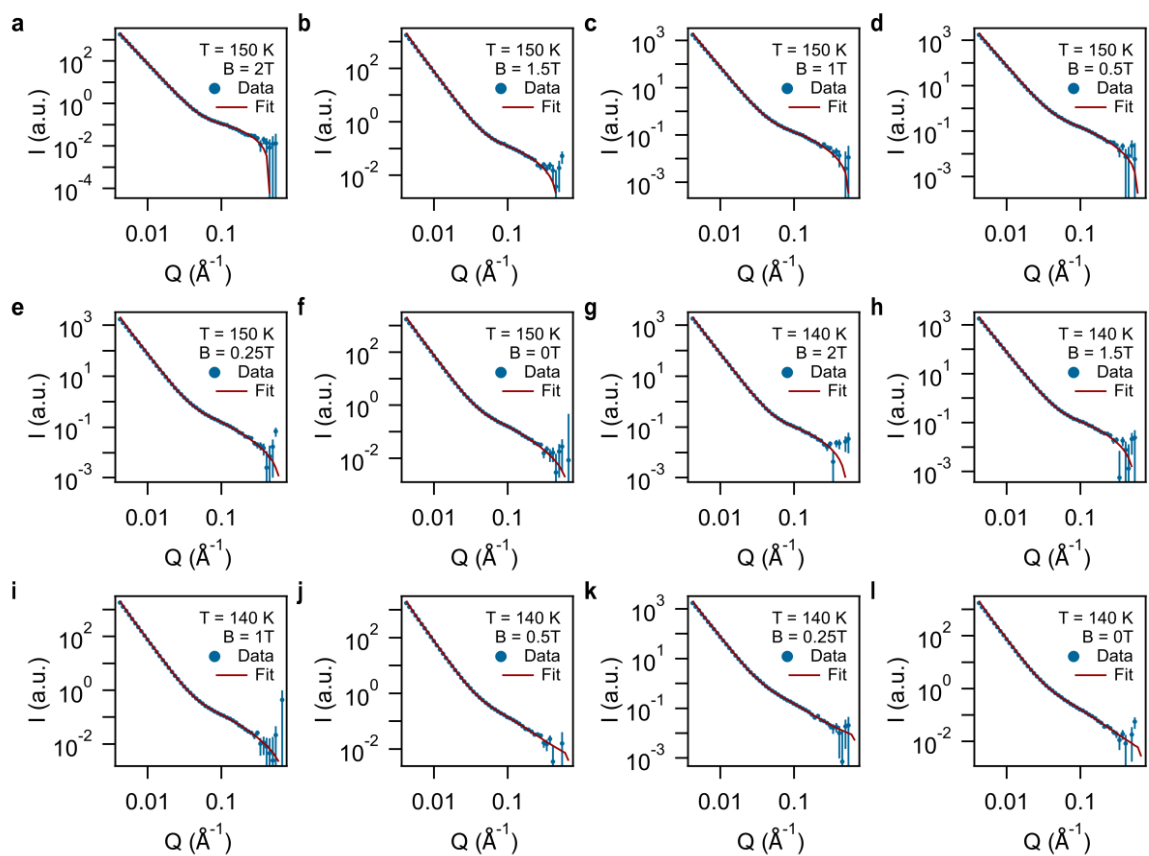

**Supplementary Figure 6.-** Fitted SANS signal at different temperatures and fields (see Methods).

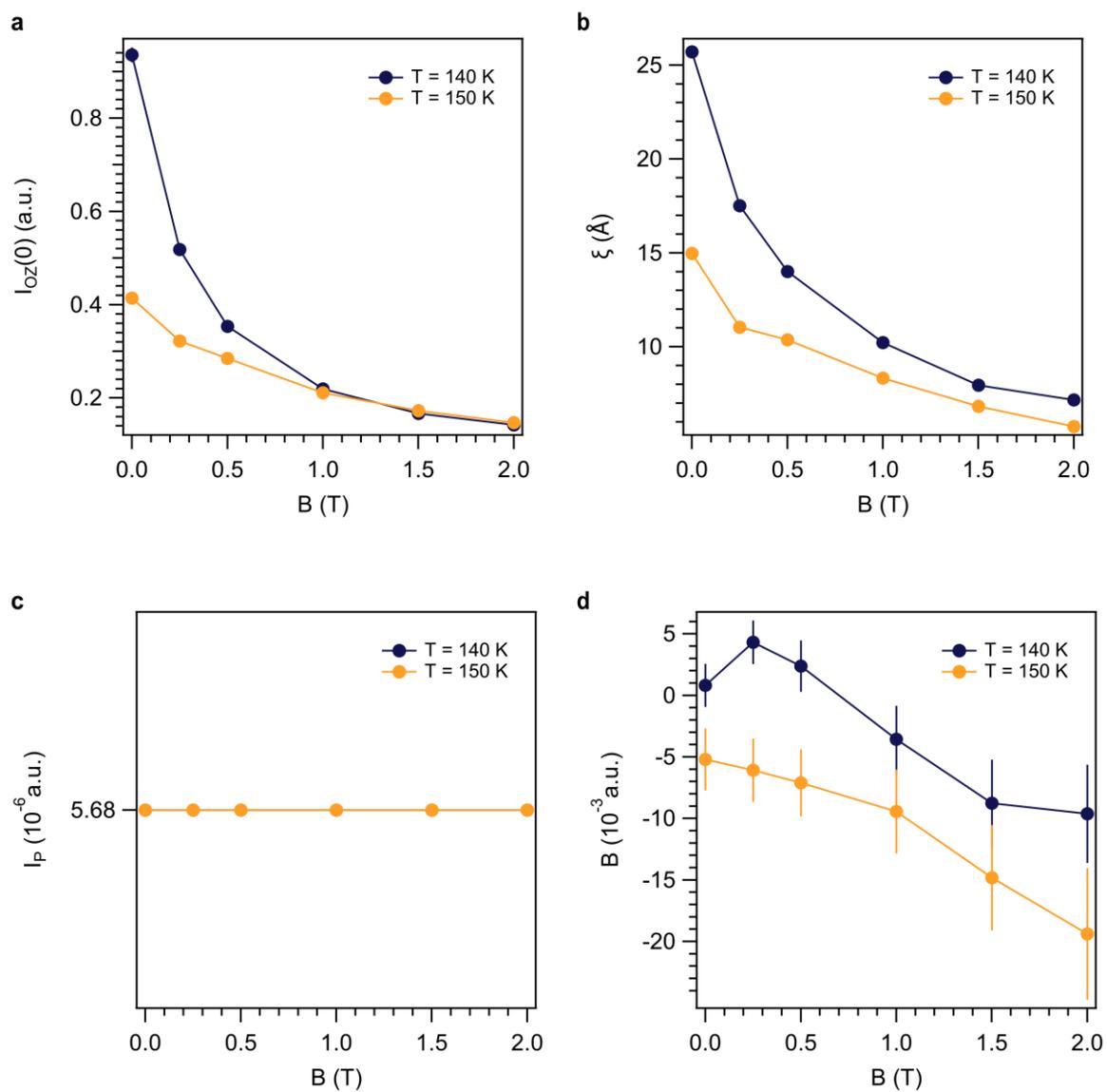

**Supplementary Figure 7.-** Thermal evolution of the intensity scale **(a)**, correlation length **(b)**, Porod scale factor **(c)** and background term **(d)** obtained from the fits shown in the **Supplementary Figure 6**.

## b. I(T) – I(300 K) analysis.

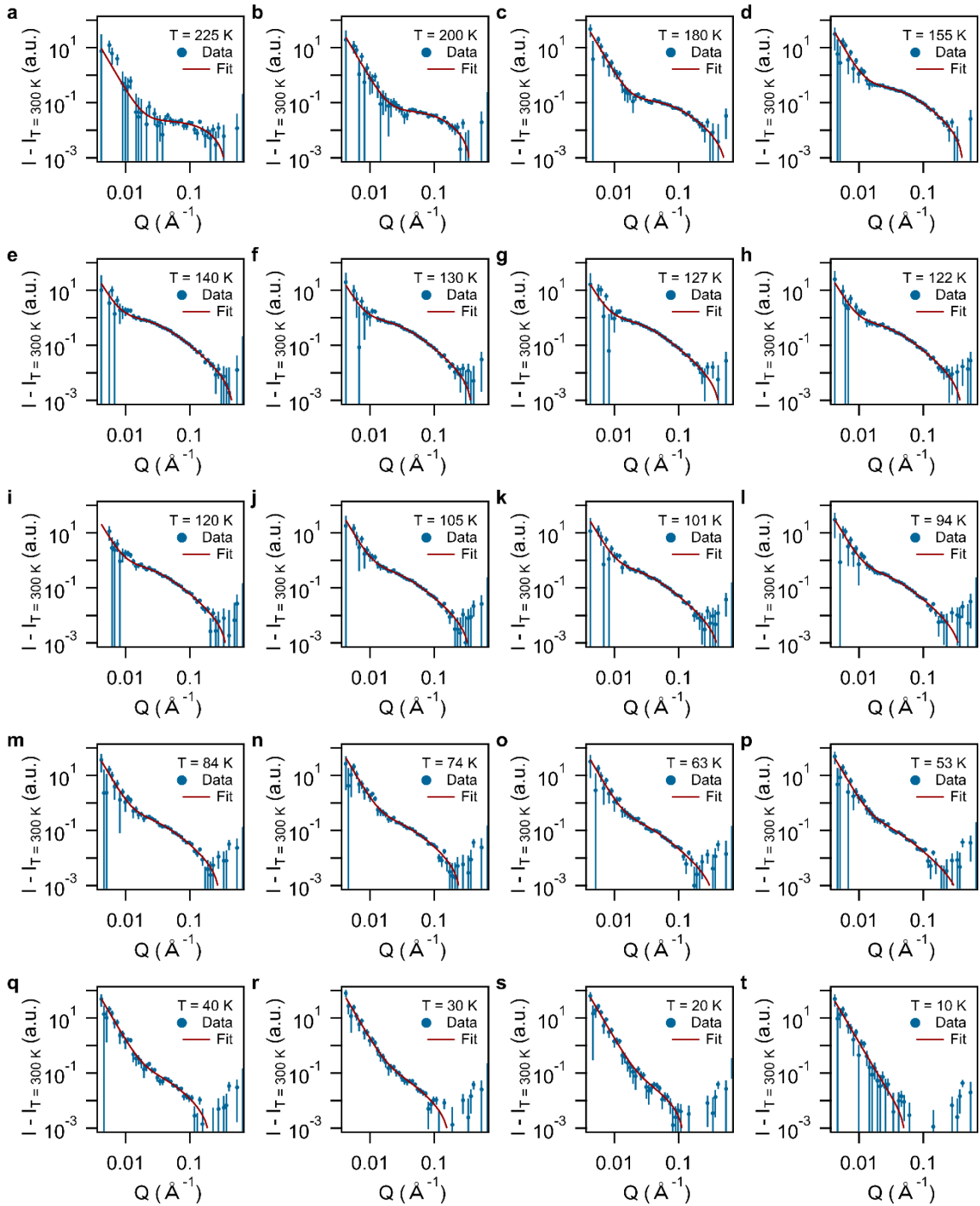

**Supplementary Figure 8.-** Magnetic contribution of the SANS signal fitted following an Ornstein-Zernike law plus a Porod term and background at different temperatures.

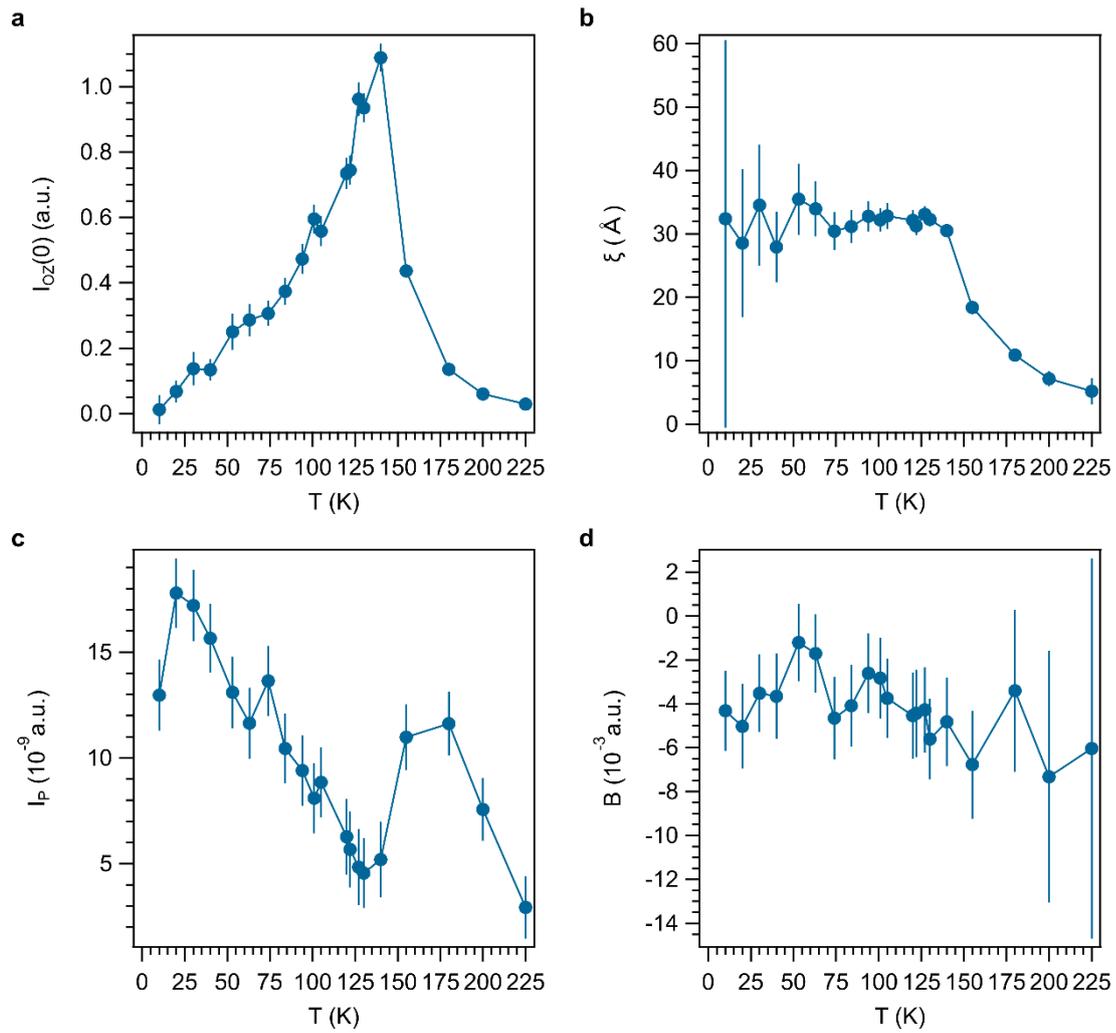

**Supplementary Figure 9.-** Thermal evolution of the intensity scale **(a)**, correlation length **(b)**, Porod scale factor **(c)** and background term **(d)** obtained from the fits shown in the **Supplementary Figure 8**.

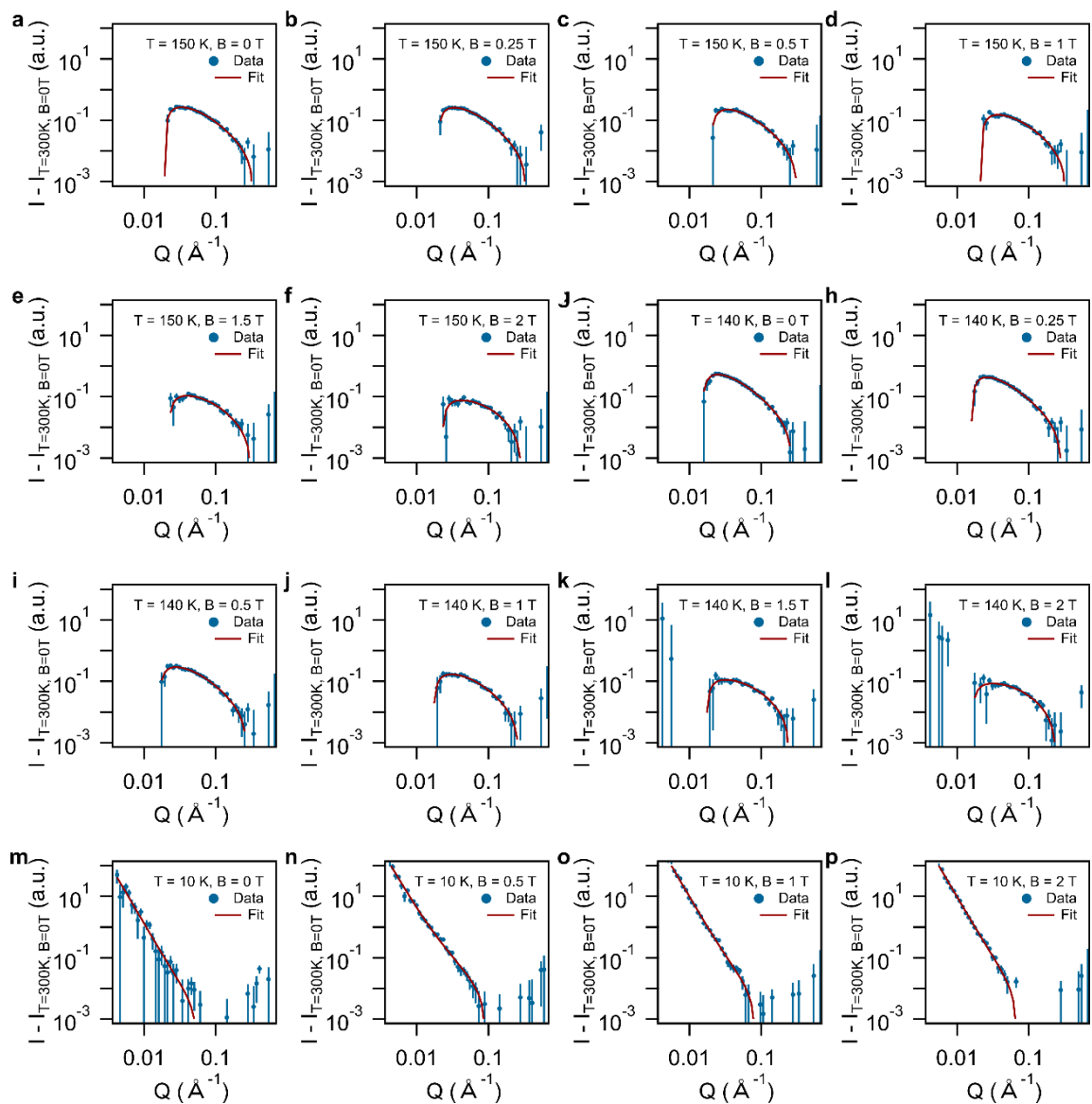

**Supplementary Figure 10.-** Magnetic contribution of the SANS signal fitted following an Ornstein-Zernike law plus a Porod term and background at different temperatures and magnetic fields.

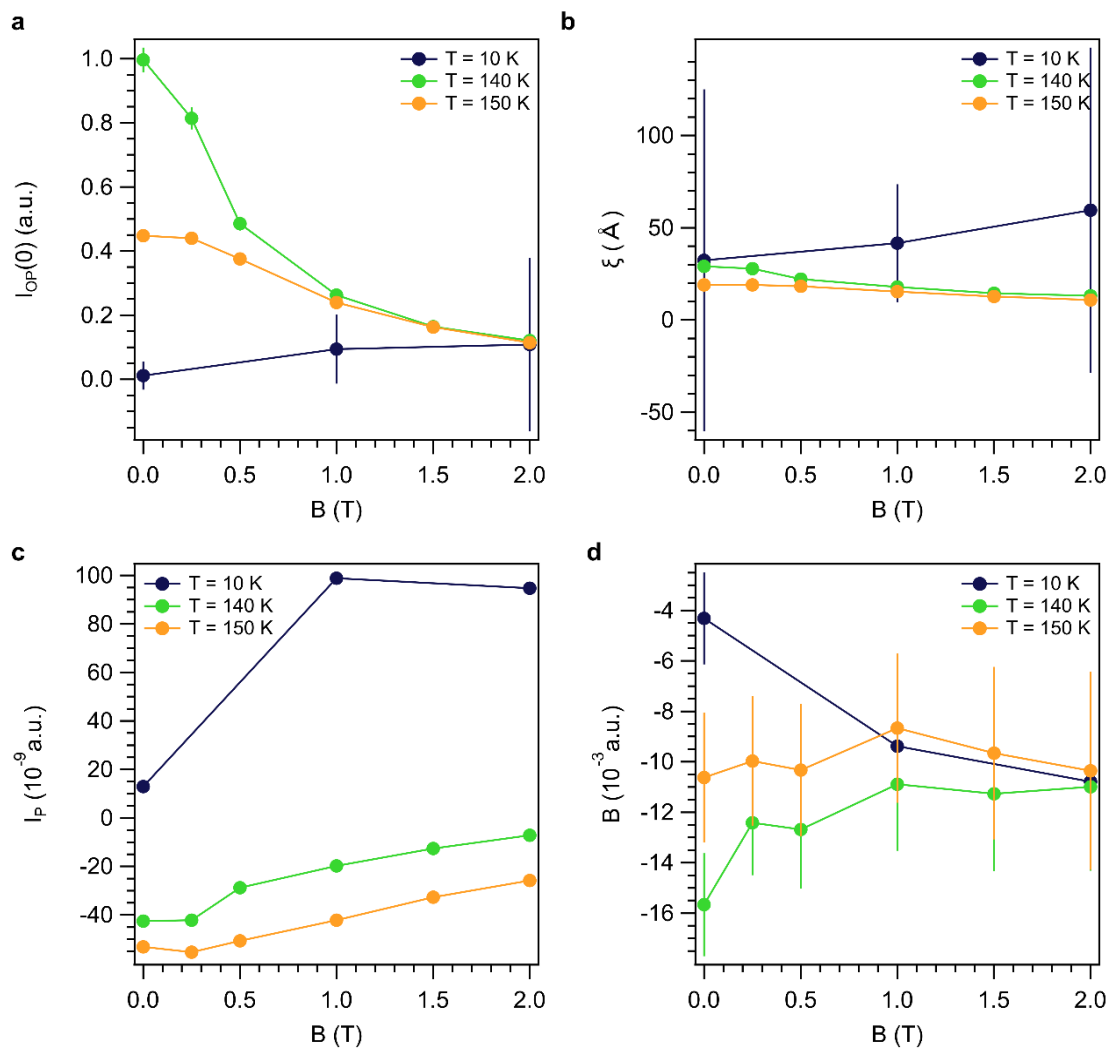

**Supplementary Figure 11.-** Field evolution of the intensity scale **(a)**, correlation length **(b)**, Porod scale factor **(c)** and background term **(d)** at different temperatures obtained from the fits shown in the **Supplementary Figure 10**.

## 3. Simulations.

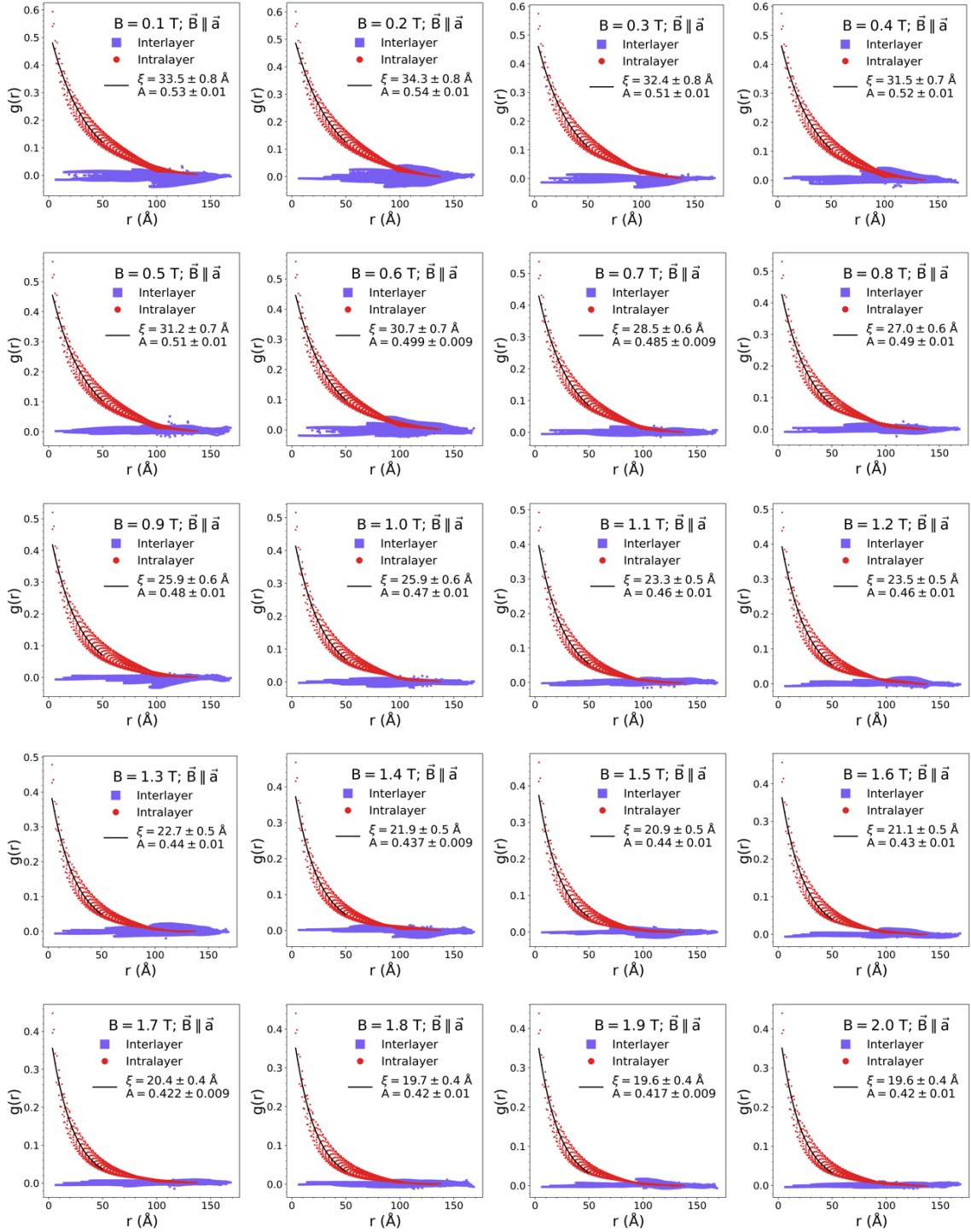

**Supplementary Figure 12.-** Distance-averaged dynamic spin-spin correlation function and exponential fit to $A \cdot e^{-r/\xi}$ for the intralayer part of the correlation function for different values of magnetic field at T = 140 K. Magnetic field is oriented along the direction of crystallographic axis *a*. The spin pairs are separated into two groups: interlayer (spin pairs where both spins belong to different layers) and intralayer (spin pairs where both spins belong to the same layer).

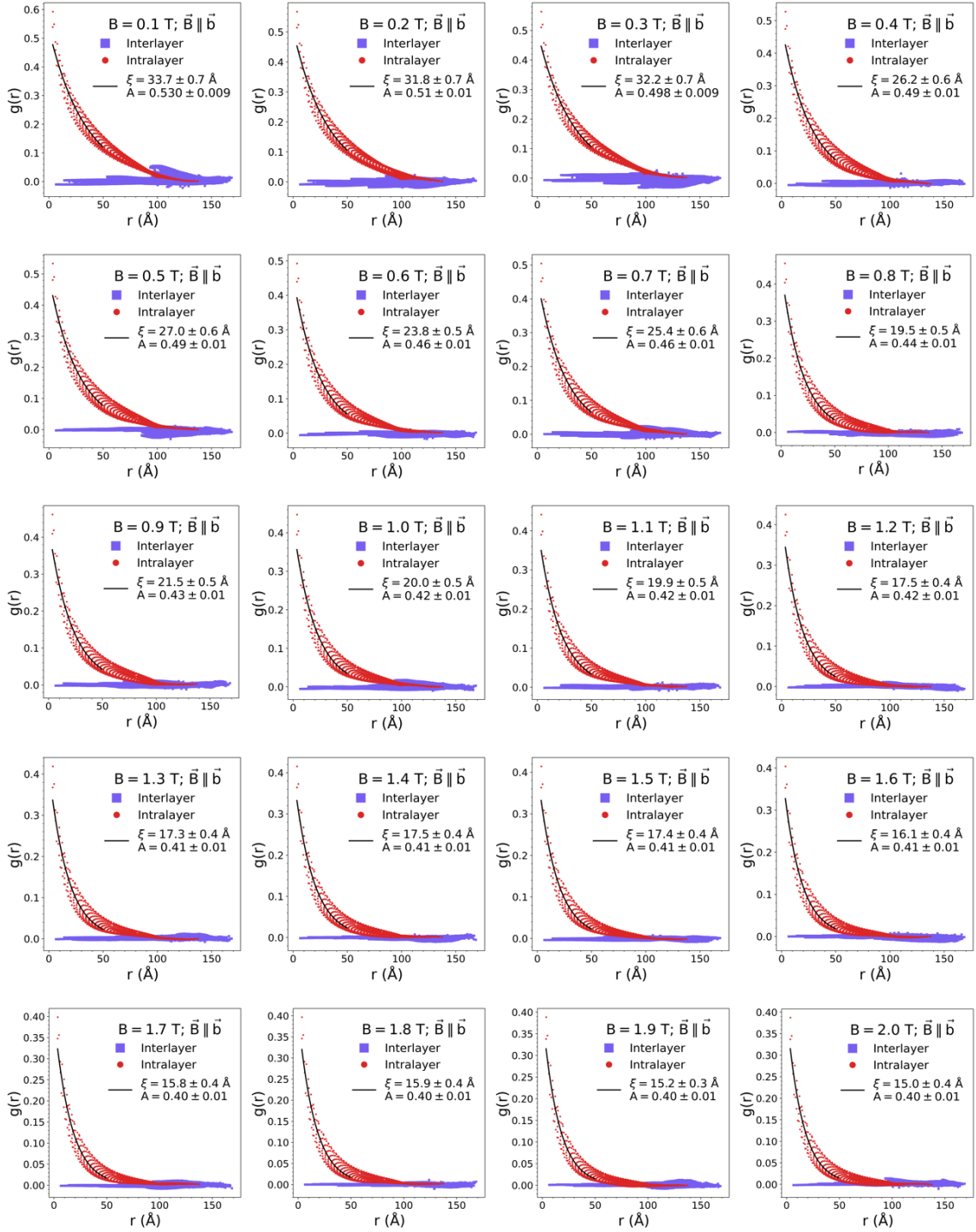

**Supplementary Figure 13.-** Distance-averaged dynamic spin-spin correlation function and exponential fit to $A \cdot e^{-r/\xi}$ for the intralayer part of the correlation function for different values of magnetic field at T = 140 K. Magnetic field is oriented along the direction of crystallographic axis *b*. The spin pairs are separated into two groups: interlayer (spin pairs where both spins belong to different layers) and intralayer (spin pairs where both spins belong to the same layer).

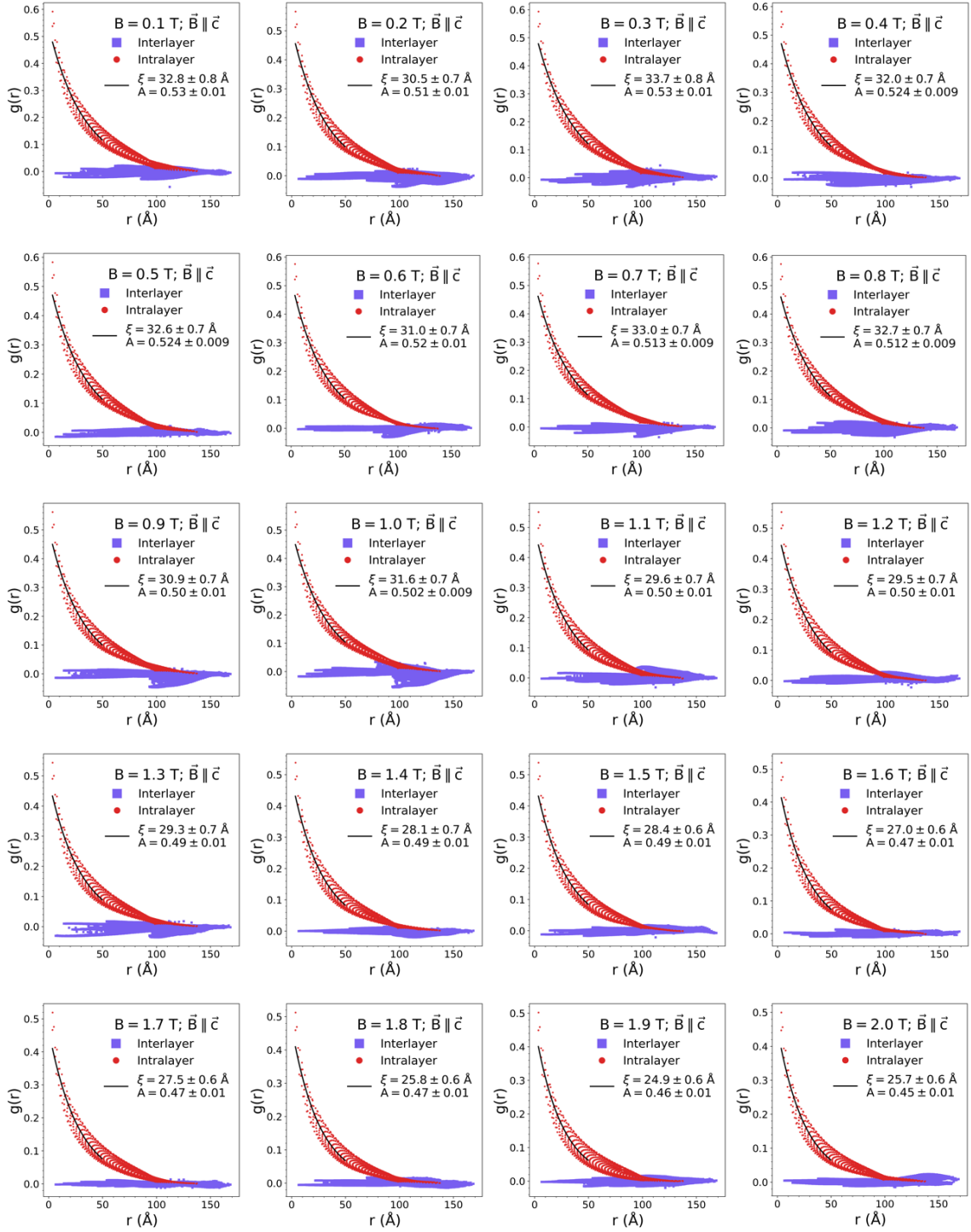

**Supplementary Figure 14.-** Distance-averaged dynamic spin-spin correlation function and exponential fit to $A \cdot e^{-r/\xi}$ for the intralayer part of the correlation function for different values of magnetic field at T = 140 K. Magnetic field is oriented along the direction of crystallographic axis *c*. The spin pairs are separated into two groups: interlayer (spin pairs where both spins belong to different layers) and intralayer (spin pairs where both spins belong to the same layer).

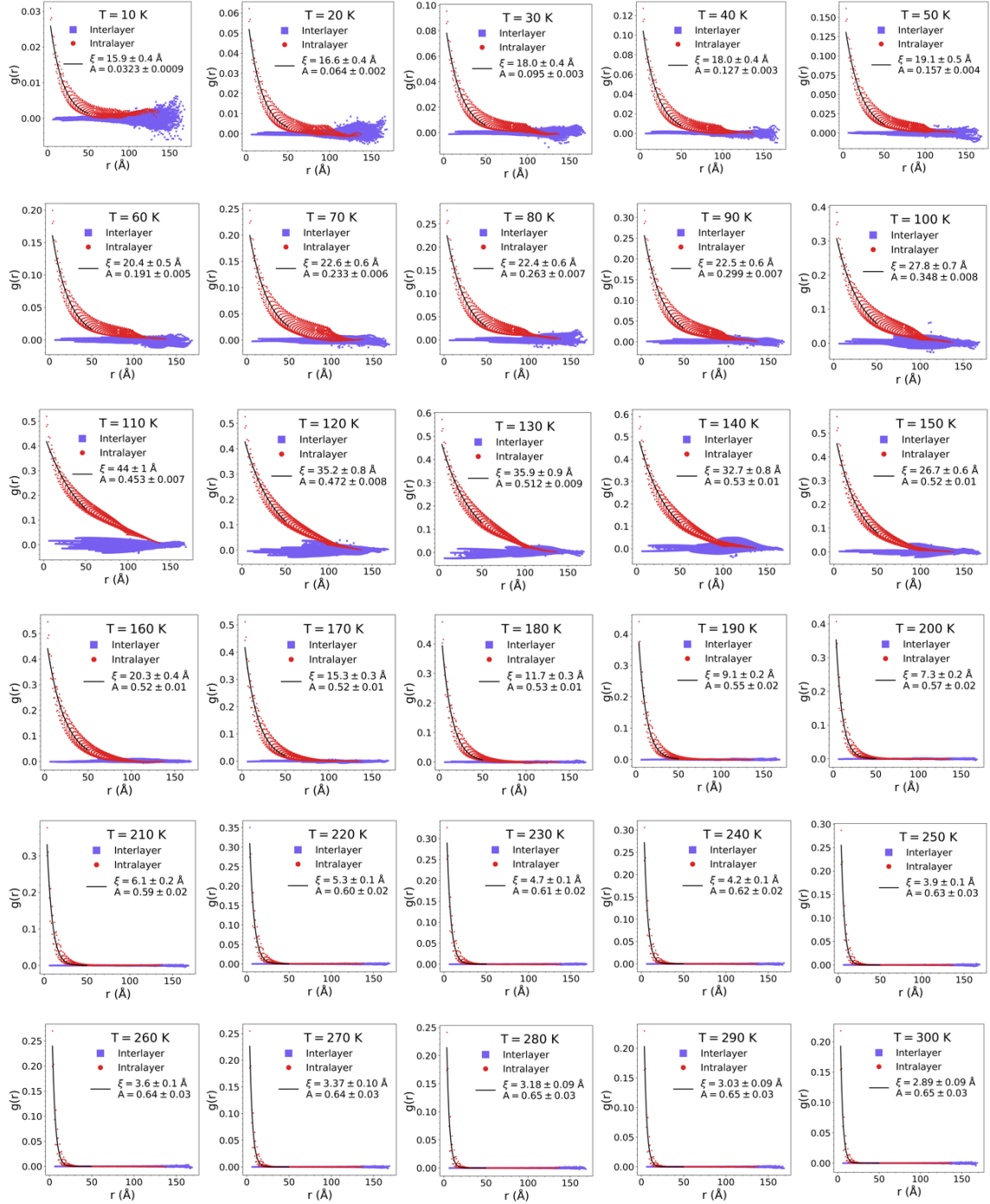

**Supplementary Figure 15.-** Distance-averaged dynamic spin-spin correlation function and exponential fit to $A \cdot e^{-r/\xi}$ for the intralayer part of the correlation function at different temperatures and zero applied magnetic field. The spin pairs are separated into two groups: interlayer (spin pairs where both spins belong to different layers) and intralayer (spin pairs where both spins belong to the same layer).

**Supplementary Table 1.-** Parametrization of the spin Hamiltonian. The notation is the one of the Hamiltonian from the main text: spins are not normalized; double counting is included. We provide unscaled parameters in this table. Actual parameters used in simulations are those multiplied by the factor 1.5.

| | Intralayer | | | | | | |
|---|---|---|---|---|---|---|---|
| Neighbor's order | 1 | 2 | 3 | 4 | 5 | 6 | 7 |
| $J_{isotropic}$, meV | -0.95 | -1.69 | -0.835 | -0.045 | -0.045 | 0.185 | -0.145 |
| DMI, meV | (0, 0.07, 0) | - | (0.18, 0, 0) | - | - | - | - |
| | Interlayer | | | | | | |
| Neighbor's order | 1 | | | 2 | | | |
| $J_{isotropic}$, meV | 0.0008 | | | -0.0008 | | | |
| | SIA anisotropy | | | | | | |
| $A_{xx}$, meV | -0.018 | | | | | | |
| $A_{yy}$, meV | -0.034 | | | | | | |
| $A_{zz}$, meV | 0.018 | | | | | | |

### 4. Elastic SANS contribution.

We note that there is certainly an inelastic contribution to our signal, as expected as well in a reactor-based source (where the intensity is integrated without analyzing the energy). Nonetheless, as the neutron's speed is smaller than the speed of sound in the solid, the phonon contribution can be considered negligible (see eq. 3.43 from A. Michels and J. Weissmüller, Rep. Prog. Phys. 71 066501, 2008). It is more challenging to assess the magnon contribution. However, we do not observe any significant variation upon different wavelengths (as expected for an inelastic contribution), thus indicating that the signal is mostly elastic (A. Michels and J. Weissmüller, Rep. Prog. Phys. 71 066501, 2008), as shown in the **Supplementary Figure 16**.

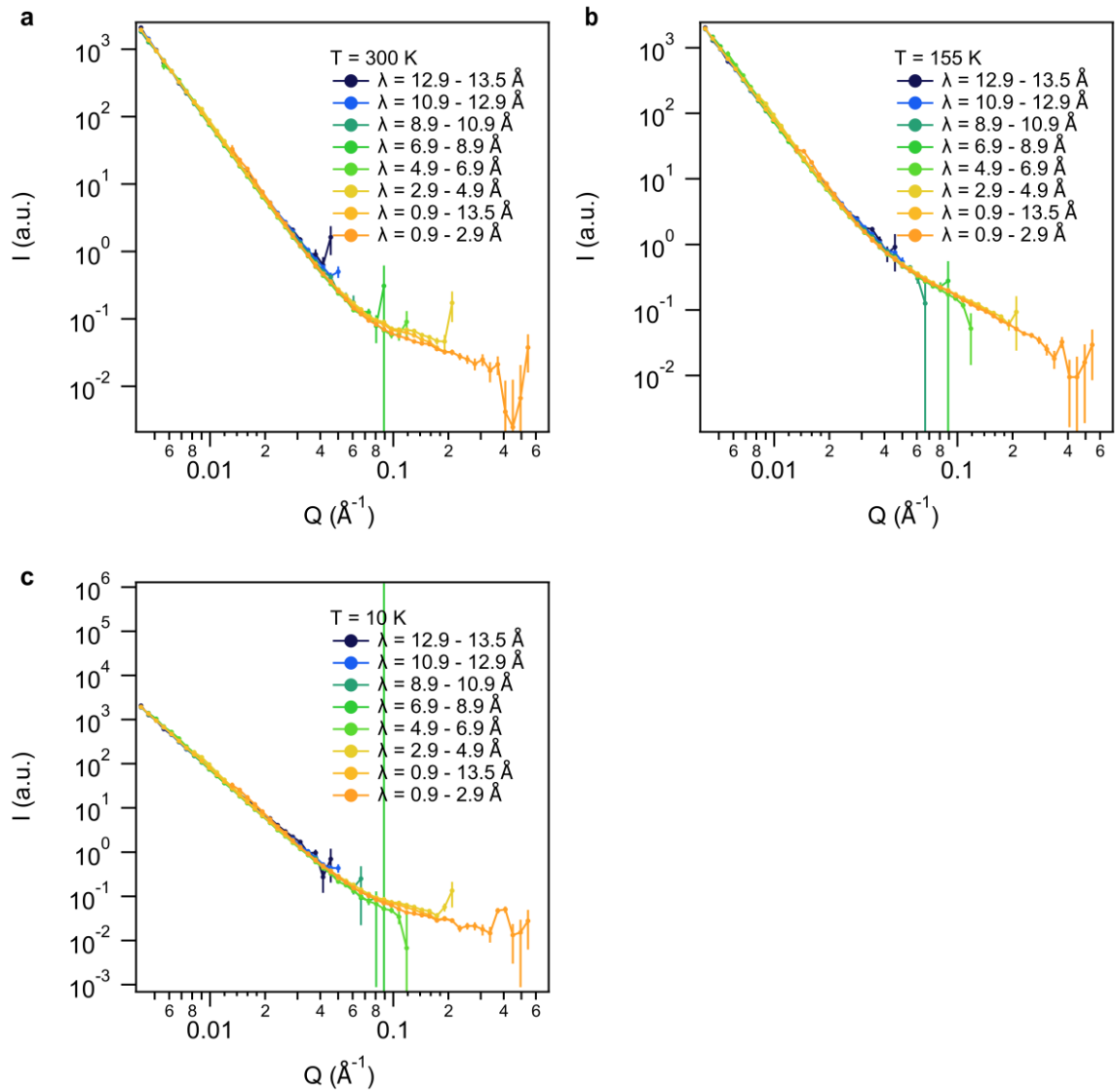

**Supplementary Figure 16.-** Selected 1D data reduction using different sections of the incoming wavelength spectrum. It can be seen that the overlap between the curves is very good, indicating the scattering intensity is only dependent on the momentum transfer of the neutrons.

20